\documentclass[aps,prb,twocolumn,showpacs,superscriptaddress,amsmath,amssymb,floatfix,10pt,reprint]{revtex4-1}
\usepackage{graphicx}  
\usepackage{dcolumn}   
\usepackage{bm}        
\usepackage{amssymb}   
\usepackage{framed}
\usepackage{amsmath}
\usepackage{multirow}
\usepackage{hhline}
\usepackage{color}
\definecolor{bluegreen}{rgb}{0,0.2,0.8}
\usepackage{subfigure,amsmath,verbatim,moreverb}
\usepackage{tabularx}
\usepackage{adjustbox}
\usepackage{lipsum}
\usepackage{longtable}
\usepackage{booktabs}
\usepackage{latexsym}
\usepackage{dcolumn}
\usepackage{amsmath}
\usepackage{epsf}
\usepackage{float}
\usepackage{hyperref}

\usepackage{etoolbox}
\AtBeginEnvironment{align}{\setcounter{subeqn}{0}}
\newcounter{subeqn} %

\begin{document}

\title{Renormalization group analysis of weakly interacting van der Waals Fermi system}

\author{Sushant Kumar Behera}
\email{sushant@niser.ac.in}
\affiliation{Density Functional Theory \& Quantum Simulations Group~(DFTQSG), School of Physical Sciences, National Institute of Science Education and Research, HBNI, Bhubaneswar 752050, India}

\author{Madhavi Ahalawat}
\email{mahlawat@as.iitr.ac.in}
\affiliation{Department of Applied Science and Engineering, Indian Institute of Technology Roorkee, Roorkee-247667, India.}

\author{Subrata Jana}
\email{subrata.jana@niser.ac.in}
\affiliation{Density Functional Theory \& Quantum Simulations Group~(DFTQSG), School of Physical Sciences, National Institute of Science Education and Research, HBNI, Bhubaneswar 752050, India}

\author{Prasanjit Samal}
\email{psamal@niser.ac.in}
\affiliation{Density Functional Theory \& Quantum Simulations Group~(DFTQSG), School of Physical Sciences, National Institute of Science Education and Research, HBNI, Bhubaneswar 752050, India}

\author{Pritam Deb}
\email{pdeb@tezu.ernet.in}
\affiliation{Advanced Functional Material Laboratory (AFML), Department of Physics, Tezpur University (Central University), Tezpur-784028, India}

\date{\today}

\begin{abstract}
Weak-coupling phenomena of the two-dimensional Hubbard model is gaining momentum as a new interesting research field due to its extraordinarily rich behavior as a function of the carrier density and model parameters. Salmhofer [{\it Commun. Math. Phys}. \textbf{194}, 249 (1998);{\it Phys. Rev. Lett}. {\bf 87}, 187004 (2001)] developed a new renormalization-group method for interacting Fermi systems and Metzner [{\it Phys. Rev. B} {\bf 61}, 7364 (2000);{\it Phys. Rev. Lett}. {\bf 85}, 5162 (2000)] implemented this renormalization group analysis of the two-dimensional Hubbard model. In this work, we demonstrate the spin-wave dependent susceptibility behavior of model graphene-phosphorene van der Waals heterostructure in the framework of renormalization group approach. We implement signlet vertex response function for the weakly interacting van der Waals Fermi system with nearest-neighbor hopping amplitudes. This analytical approach is further correlated with {\it ab initio} simulation results and extended for spin-wave dependent susceptibility behavior with possible experimental protocols. We present the resulting compressibility and phase diagram in the vicinity of half-filling, and also results for the density dependence of the critical energy scale.

\end{abstract}


\maketitle

\section{Introduction}
Hubbard model \cite{1,2,3,5,PhysRevLett.111.266801,PhysRevB.99.205150,PhysRevX.10.021042} becomes an effective tool to describe the many-body correlation effects in condensed matter physics~\cite{PhysRevResearch.2.033087,PhysRevLett.122.086402,PhysRevLett.121.026402,PhysRevX.8.031089}. It exhibits various different channels for the transition of normal metallic behavior by an independent tunability of the electronic band structure \textit{via} a single tight-binding parameter and the half-filling nature on the hexagonal lattice bilayer system \cite{PhysRevResearch.2.033087,Cao2018,PhysRevLett.121.026402}. In this aspect, weak-coupling phenomena of the two-dimensional (2D) Hubbard model is gaining momentum as a new interesting research field because of its extraordinarily rich behavior as a function of the carrier density and model parameters~\cite{PhysRevLett.121.026402}. Low-dimensional interacting lattice electron systems are generally featured with Fermi liquid instabilities driven by long-lived electron-electron or electron-hole excitations near the Fermi surface. These excitations lead to various types of long range correlations and destruction of Fermi liquid behavior. 

However, for low dimensional Fermi systems, we recall the Renormalization Group (RG) is the most promising and best controlled analytical approach~\cite{Salmhofer1998,PhysRevB.82.155404,PhysRevB.88.241401,Jongeward1983}. In 2D or bulk systems, RG methods are probably the least biased approach to study such systems at weak coupling and low energy scales in a transparent way~\cite{30,PhysRevB.61.13609,PhysRevB.61.7364}. This has led to the development of alternative flow schemes~\cite{PhysRevLett.85.5162,PhysRevLett.87.187004}. They provide the possibility to tackle the competing instabilities from different angles. Fortunately, it turns out that besides well-understood conceptual differences the flow of these schemes agrees quite favorably over a wider parameter range of models like the 2D Hubbard model. This connection of RG method and 2D Hubbard model encourages to further proceed on a firm approach to understand the instabilities in 2D system. In this regard, van der Waals (vdW) heterostructure systems~\cite{PhysRevB.102.085103} are coming under direct focus, where RG Theory and 2D Hubbard model can be implemented for practical realization to explore the concept of spin-wave assisted device application~\cite{D0CP00836B}. 

To realize and analyze prototypic 2D model for the electronic degrees of freedom, Hubbard model including RG theory is promising in case of 2D planes \cite{26,PhysRevLett.116.076803}. In particular, 2D Hubbard model~\cite{PhysRevB.82.075424} is appropriate to explain interesting phenomena of weakly interacting systems which can be tuned by their carrier density. Besides, large scale numerical algorithms are worked out to solve such 2D interacting Hubbard model including related model parameters \cite{27}. In such systems, weak interaction reveals antiferromagnetic ground state close to half-filling and symmetric \textit{s}-wave phase away from half-filling region. In such cases, conventional perturbation theory fails at a certain region close to the half-filling region, where challenging and frequent infrared deviations raise due to weakly interacting Fermion system \cite{29,30,Naji2013,PhysRevLett.116.076803} and \textit{van Hove singularities}~\cite{PhysRevB.92.085423}, non-smooth critical points in the density of states (DOS) of solids in their Brillouin zone, \cite{31} including the 2D monolayer sheet system \cite{32}. Moreover, appropriate model for 2D weakly interacting vdW Fermi systems with singularities within weakly interacting region is still lacking in the current analytical prospects. To overcome this fact, RG theory is presently the most efficient and promising analytical technique for such 2D weakly interacting Fermi systems \cite{35,36,D0CP00836B,C9CP05252F,PhysRevLett.116.076803}. 2D Hubbard model is the simplest of a class of models describing electrons moving on a lattice having weakly-interacting electron-electron interactions. However, electron-electron interactions can break one or more symmetries of the Hamiltonian, and this leads to phases such as anti-ferromagnets, charge or density of spin-waves, etc. Besides, RG theory is presently the most promising and best controlled approach to low dimensional systems with competing singularities at weak coupling region. In this regard, direct device application through RG theory is unrealistic in pristine form, until and unless corroborated with vdWs heterolayer system using {\it ab initio} pseudopotential method. Thus, it is preferable to implement RG theory approach, unlike Dirac-like field theory to study the stability of the system supported by pseudopotential based density functional theory (DFT) simulation, unlike full potential basis of tight-binding method for practical realization supporting experimental scope.

In this work, we implement RG theory to the 2D Hubbard model of Graphene-Phosphorene (Grp-P) heterostructure considering nearest neighbor hopping amplitudes of dense electron distribution profile close to half-filling region. Here, we particularly calculate the trend of collective two particle couplings in a single loop stage, completely ignoring the inappropriate energy and linear momentum dependency, but considering vital tangential momentum dependency being a dynamic parameter. We further extend the RG theory to obtain the susceptibility values to analyze the physical response of uncertainties gestured by diverging couplings in weakly interacting vdW heterostructure. Charge-density screening, spin-density and singlet function based susceptibilities have been considered during analytical calculations at various pairing symmetries. Here, spin-density phase diagram has been presented close to half-filling region of this 2D Hubbard model. The obtained results show the critical energy scale behavior away from the half-filling region with perfect nesting and room temperature stability of the designed system for further functionality. 

\section{Renormalization Group Equations}
In this section we revisit Salmhofer’s renormalization group approach~\cite{Salmhofer1998} for general weakly interacting Fermi systems with the explicit flow equations for effective two particle interactions on a one-loop level, and finally derive one-loop flow equations for several physical observables, like compressibility, spontaneous diamagnetic free susceptibility, etc., which will later be used for our stability analysis of the 2D Hubbard model.
\subsection{Presentation of Functional Integral form}
We have taken weakly interacting spin-half Fermions of single-particle basis state with crystal momentum ($p_{c}$), {\it up} or {\it down} spin-wave projection, $\sigma\epsilon[{\uparrow,\downarrow}]$ and kinetic energy ($E_{q}$). Thus, the behavioral aspects of the system can be determined via following Eq.~\ref{eq_2},
\begin{equation}
 S[\chi,{\chi}']=\sum_{Q}(if_{0}-{\chi}_{q}){\chi}_{Q}'{\chi}_{Q}-\zeta[\chi,{\chi}']
 \label{eq_2}
\end{equation}
where, Q=($f_{0}$,p,$\sigma$) presents a multi index parameter connecting the frequency ($f_{0}$) with each particle quantum numbers; Grassmann variables (${\chi}_{Q}$ and ${\chi}_{Q}'$) are allied to the formation and extinction operators, ${\chi}_{Q}$=${\epsilon}_{Q}$-$\mu$ is the energy related to a particle depending on the chemical potential ($\mu$) and $\zeta[\chi,{\chi}']$ is a random many body coupling factor \citep{40}. \\
In the calculation steps, all the connected and dependent transient flow equations have been acheived from the functional via implimenting Green's functions with the normalized Gaussian measure (given below in Eq.~\ref{eq_3}).
\begin{equation}
 G[\eta,{\eta}']=log{\int {\partial}{\mu}_{c}[\chi,{\chi}']exp^{-\zeta[\chi,{\chi}']}exp^{({\chi}',\eta)+\chi,{\eta}'}}
 \label{eq_3}
\end{equation}

Green's function is an efficient and yet accurate tool to predict the variables in flow equations and maintain proper correlation among the RG theory equations. The infinite hierarchy in flow equations is essential to check at a single-loop level to predict dominant structural uncertainties in the vdW system within weakly interacting limit, neglecting other irrelevant components of the effective interactions (details shown in Supplimentary Material~\cite{SM}). 

\begin{figure}[th!]     
\centering           
\includegraphics[width=7.5cm,height=5cm]{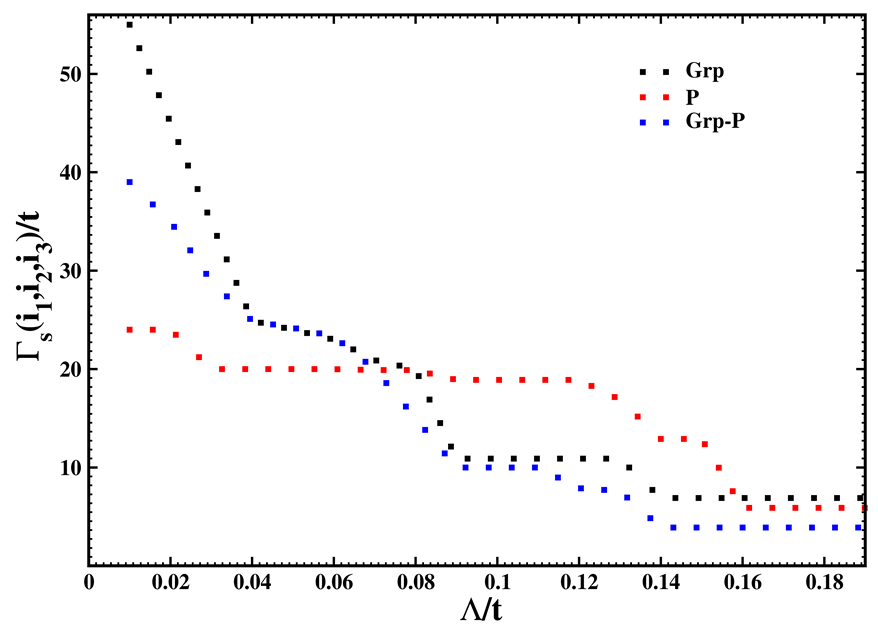}
\caption{Flow of the singlet vertex function, $\Gamma$, as a function of $\varLambda$ for several selections of momenta values, {\it kF$_1$}, {\it kF$_2$} and {\it kF$_3$} for the {\it Grp-P} heterostructure system with a comparison to the pristine graphene and phosphorene sheet. The model parameters are U=t and t=0, and the chemical potential $|\mu|$=0.005.}
  \label{fig_1}
\end{figure}
\subsection{Vertex Response Function}
Renormalization Group method is presently the most promising and best controlled analytical approach to low dimensional Fermi systems with competing singularities at weak coupling region. Renormalization groups are commonly used as the existence of power-law singularities near critical points and half-filling regions. Here, the classic predictions are sometimes disturbed and logarithmic and exponential corrections are required in step-by-step process to address van der Waals low dimensional systems. To formulate the mechanism of Renormalization group theory is an interesting aspect to systematically explain the families of weakly-interacting Fermi systems united by common scaling variables. In the weak coupling case, the low-energy physics can be obtained via naive Hartree-Fock perturbation theory with weak coupling RG analysis. 

Moreover, energy corrections of each iteration step are confined to one-particle vertex function ($\Gamma$) and used for further calculations. The spin-wave structure of $\Gamma$ can be used for a spin rotation invariant system with response flow equation, which is given as follow in Eq.~\ref{eq_4},
\begin{widetext}
\begin{equation}
 \Gamma(K_{1}',K_{2}';K_{1},K_{2})={\Gamma}_{s}(K_{1}',K_{2}';K_{1},K_{2}){\varLambda}^{1/2}S_{{\sigma}_{1}',{\sigma}_{2}';{\sigma}_{1},{\sigma}_{2}}+
 {\Gamma}_{t}(K_{1}',K_{2}';K_{1},K_{2}){\varLambda}^{3/2}T_{{\sigma}_{1}',{\sigma}_{2}';{\sigma}_{1},{\sigma}_{2}}
  \label{eq_4}
\end{equation}
\end{widetext}
where, $K_{1}$ and $K_{2}$ presents the multi indexed particle quantum numbers for both spin arrangements, singlet {\it S} and triplet {\it T} projection operators are used for two particle spin space, $\varLambda$ is the energy scale varying at a functional trend for singlet and triplet operators. 
\subsection{Free susceptibility}
We consider the spontaneous response to a field perpendicular to the plane. Here, the susceptibility follows like Eq.~\ref{eq_6},
\begin{equation}
 \chi(q)={\xi}^{-1}{\langle{\Phi(q)}{\Phi'(q)}\rangle}_{F_{E}=0}
 \label{eq_6}
\end{equation}
Meanwhile, translation and spin rotation invariant systems are considered with conserved charge in the absence of an external field ($F_{E}$). As a consequence, in such normal and non-symmetrical broken phase, $\langle{\Phi(q)}\rangle$, the expectation value nullifies at $F_{E}(q)\rightarrow 0$. 
\subsection{Compressibility}
In case of the density $D = [n]$, the compressibility $C = \partial [n] / \partial \mu_D$. 
The scaling relation then implies Eq.~\ref{eq_ap_11}
\begin{equation}
\label{eq_ap_11}
C(T) = b^{x} C (b^{1-x} T^Y), 
\end{equation}
where, {\em x} stands for a constant quantity, which leads to Eq.~\ref{eq_ap_12}
\begin{equation}
\label{eq_ap_12}
C(T) \propto T^{1/2x+Y}, 
\end{equation}
with Eq.~\ref{eq_ap_13}
\begin{equation}
\label{eq_ap_13}
x = \frac{0.5}{Y} 
\end{equation} 
This is used for the weakly interacting systems to evaluate the compressibility factor.
\section{Application to 2D Hubbard Model}
\subsection{Linear Response Behaviour}
We extend the RG equations to obtain linear response behaviour ({\it i.e.} spin-wave susceptibility value) of the system towards finite applied external electric field. Finite external field ($F_{E}$) leads to an additional contribution to the action of the flow equation, which is given as follow in Eq.~\ref{eq_5}
\begin{equation}
 \zeta[F_{E};\xi,{\xi}']=-\sum_{q}[F_{E}'(q)\Phi(q)+F_{E}(q){\Phi}'(q)]
 \label{eq_5}
\end{equation}

Thus, the Grassmann variables in bilinear form (${\Phi}(q)$), Hermitian conjugate (${\Phi}'(q)$) and ${F_{E}}'(q)$ is the complex conjugate of ${F_{E}}(q)$. One can calculate the response of the field interaction towards electronic charge density, $\rho(q)$=$\sum_{k,\sigma}$(${\xi}_{k-q,\sigma}'{\xi}_{k,\sigma}$) and z-direction dependent spin density, $s^{z}$(q)=$\sum_{k}$[${\xi}_{k-q,\uparrow}'{\xi}_{k,\uparrow}$-${\xi}_{k-q,\downarrow}'{\xi}_{k,\downarrow}$] and response to coupling fields which interact with the change in singlet operator, where $\partial(k)$ is an even parity function satisfying the orbital symmetry of the coupling operator ({\it i.e.} s wave). The linear response of the expectation value with the flow equation of the partition function. 
\subsection{Weakly interacting Fermi system}
We  study the effects of weak interaction in this 2D vdW system by RG theory analysis. It is obvious that over a wide range of energies the screened interaction is a relevant perturbation, and the system fails in the strong coupling regime displaying non-Fermi liquid behavior with power law energy dependence of the quasiparticle residue (discussed in the Supplimentary Material \cite{SM}in details). We describe a semimetal-semiconductor transition and associated quantum criticality in a 2D vdW system with particular emphasis on graphene monolayer sheet\cite{RevModPhys.81.109,PhysRevB.77.041409,PhysRevB.77.195413}. Moreover, the system enters into a weak coupling regime below a certain energy range. In this regime, only logarithmic renormalizations are possible\cite{PhysRevB.89.201110,PhysRevLett.99.226803,PhysRevLett.107.196803,PhysRevLett.108.046602,PhysRevB.86.165127,PhysRevLett.116.076803}. As a consequence, all the possible calculations have been performed for physical observables, like compressibility, spontaneous free susceptibility, etc.\cite{PhysRevLett.99.226803,PhysRevB.87.205138,PhysRevB.80.193411}. Other physical observables, like free susceptibility, phase transition, etc. are discussed in the Supplimentary Material~\cite{SM} in details. 
\section{VERTEX FUNCTION FINDINGS}
\subsection{Vertex response function corrections}
One can verify that higher-order corrections contain higher powers of the upper limit~\cite{HUR20091452}. The presence of anomalous dimensions in the RG theory proves that physical observables consdiered here int he work, such as spontaneous susceptibility, compressibility, etc., show indirect dependencies on response function. This strong coupling results differ quantitatively, but not qualitatively from the case of pristine graphene.

The flow of the vertex function ($\Gamma$) (Eq.~\ref{eq_4}) and the free susceptibility ($\chi_F$) (Eq.~\ref{eq_6}) values have been calculated analytically using RG theory for several choices of the bare interaction, termed as field potential (U$>$0), the next nearest neighbor hopping amplitude (t$<$0) and the chemical potential ($\mu$), where t and $\mu$ have been fixed so that the Fermi surface and particle density are approaching to $\epsilon$k and half-filling (i.e. n$\approx$1, $|\mu|\rightarrow$0), respectively. Momentum dependent flow of the vertex functions are obtained for the three considered systems at low dimensional scale. It is observed that the flow of the vertex function achieves a density value (n$\approx$0.992), {\it i.e.} close to half-filling region at t=0 and $|\mu|$=0.005 (plotted in Fig.~\ref{fig_1}). Here, the singlet part of the vertex function is taken at specified values of momenta on the Fermi surface including the values of momenta, where strong renormalization of $\Gamma$ occurs. Most probably, the singlet vertex function possesses its leading values supporting {\it umklapp} scattering along the diagonal direction of the {\it k}-space Brillouin zone. It is observed from the RG theory calculation that the density decreases away from half-filling ({\it i.e.} n$\approx$1, $|\mu|\rightarrow$0) region in a step-like manner and the calculations fail into a region, where {\it s}-wave symmetry dominates with pairing correlations at necessarily low scale energy levels indicating quantum confinement ({\it i.e.} nanostructuring) in the {\it Grp-P} heterostructure system. As a result, differential step like behaviour is prominent from the plots in Fig.~\ref{fig_1}.

\begin{figure}[th!]     
\centering           
\includegraphics[width=7.5cm,height=5.5cm]{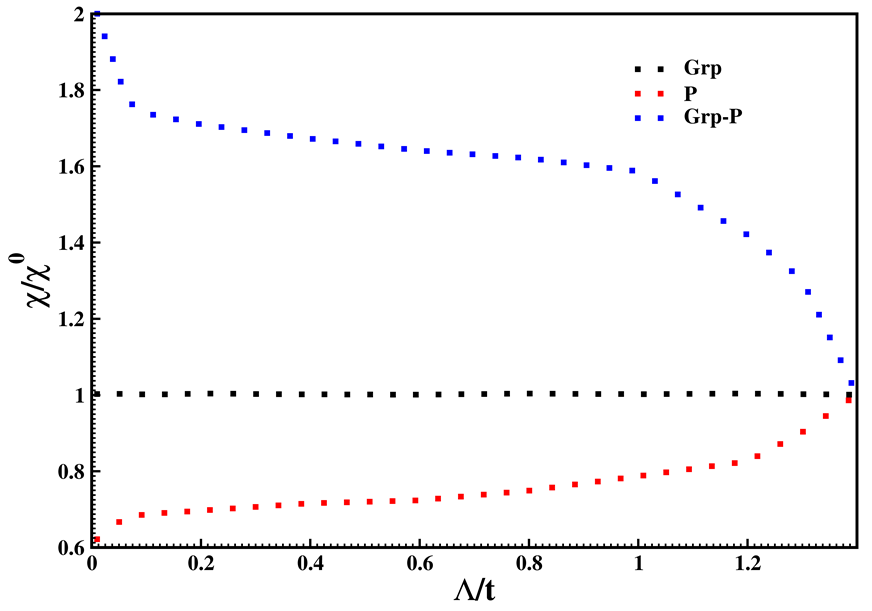}
\caption{The flow of the ratio of weakly interacting and free susceptibility values depending on $\varLambda$ values for the {\it Grp-P} heterostructure system with a comparison to the pristine graphene and phosphorene sheet. The model parameters are U=t and t=0, and the chemical potential $|\mu|$=0.005.}
   \label{fig_2}
\end{figure}

\subsection{Spin-wave dependent susceptibility behaviour}
Two types of antiferromagnetic spin-wave ({\it i.e.} incommensurate and commensurate) susceptibilities have been obtained from the numerical calculations to validate the physical instability associated with the diverging vertex function as a consequence of response functions \citep{42}. It is observed that the incommensurate spin susceptibilities can be neglected compared to their respective commensurate counterparts (shown in Fig.~\ref{fig_2}) due to the incommensurability parameter ($\delta \rightarrow$0) nearing to half-filling (n$\approx$1, $|\mu|\rightarrow$0) as a function of energy scale. Moreover, the ratios between the two susceptibility values of monolayer system in case of weakly interacting {\it s}-wave region are in the same scale, whereas the ratios of free susceptibility are much reduced indicating the validity of response functions for 2D weakly coupled Fermi systems. In addition, the ratio in case of hetero bilayer is much greater compared to its monolayer system. Thus, antiferromagnetic wave vector based spin-wave susceptibility dominates over coupled susceptibilities at low scale energy level ({\it i.e.} {\it ground state}), confirming the existence of antiferromagnetic ground state with rotational spin invariance. \\ 

\subsection{Spin-wave dependent compressibility behaviour}
\begin{figure}[th!]     
\centering           
\includegraphics[width=8.5cm,height=4.0cm]{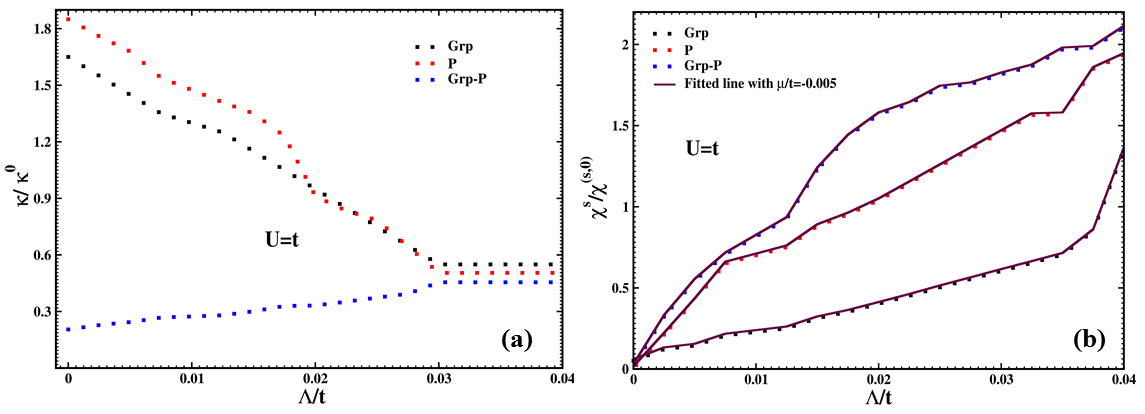}
\caption{The flow of (a) compressibility, $\kappa$, and (b) homogeneous spin susceptibility as a function of $\varLambda$ at fixed value of $|\mu|$=0.005 for U=t and n=1 for the {\it Grp-P} heterostructure system with a comparison to the pristine graphene and phosphorene sheet.}
    \label{fig_3}
\end{figure}

The values of compressibility~(Eqn.~\ref{eq_ap_11}) and homogeneous spin susceptibility have been calculated depending on the energy scale (shown in Fig.~\ref{fig_3}(a) and Fig.~\ref{fig_3}(b)), respectively for the three systems. These two parameters can be realized directly from the forward scattering of vertex function. The free compressibility, {\it $\kappa_0$}, and spin-wave susceptibility, $\chi$(s,0), are defined ignoring the probable infrared cut-off energy values. As a result, the flow equations of the vertex response function is entirely checked by the flow of the Landau functions at the initial stage of the Hubbard model at critical energy scale value, $\varLambda$=$\varLambda_{0}$. 

The compressibility is inhibited at low scale energy levels close to half-filling region, which is expected for systems having a finite bandgap near to $\mu$, with dominant uncertainty in spin-wave density. Besides, the values of the compressibility diverge with respect to the reduction in homogeneous spin-wave susceptibility values away from half-filling region, where {\it s}-wave uncertainties play major role indicating a finite spin-wave bandgap opening trend in any singlet or triplet spin-wave 2D Fermi systems. It is found that the spin-wave susceptibility flows from zero to negative values which implies that our single loop calculation fails, when it reaches strongly interacting regime. This trend is validated only in weakly interacting region applied for 2D Fermi system. Thus, one diverging compressibility is found mentioning a strong tendency toward phase transition and separation. As a result, the increase in $\kappa$ is very close to the uncertainty point, where the renormalized interactions achieve large value that the single loop findings are not further validated. The stability is prominent at higher energy scales ({\it i.e.} 0.04$\leq${$\dfrac{\varLambda}{t}$}$\geq$0.03) from the plot in Fig.~\ref{fig_3}(a) and Fig.~\ref{fig_3}(b).  Moreover, we have shown the trend of free susceptibilities ({$\chi$}$^{0}$) at specific choice of {\it t} and $\mu$ values (shown in Supplimentary Fig. S1 \cite{SM}), strongly supporting the trend of homogeneous spin-wave susceptibility. 
\subsection{Spin-wave density dependent phase diagram}
\begin{figure}
\centering
\includegraphics[width=7.5cm,height=5.5cm]{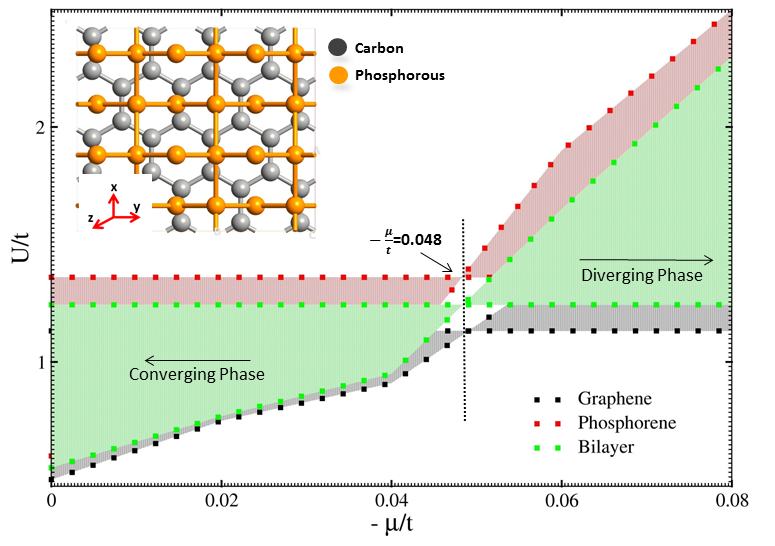}
\caption{
Spin-wave density dependent phase diagram for ($\mu$, U) phase diagram for t=0 close to half-filling (n$\approx$1, $|\mu|\rightarrow$0) region; the solid line represents the spin-density-wave regime of the {\it Grp-P} heterostructure system with a comparison to the pristine graphene and phosphorene sheet. Inset is showing the heterostructure system. 
}\label{fig_4}
\end{figure}

We show the phase diagram of $\mu$ and {\it U} at half-filling measured by the prevailing uncertainty from the flow equation (shown in Fig.~\ref{fig_4}). The leading uncertainty region in case of {\it commensurate} spin-wave density is detached from the {\it s}-wave coupling region {\it via} a narrow strip where {\it incommensurate} density fluctuations show dominant effect with {\it q}=($\pi$, $\pi$-$\delta$). For small value of {\it U}, the region surrounding half-filling is exponentially small, where uncertainties in spin-wave density play a dominant effect. The value of ({\it t, U}), {\it in-plane} phase diagram is shown in Fig.~\ref{fig_4} with $|\mu|$=4t and $\mu${$<$}0 (below half-filling). It is observed that the chemical potential is fixed at the {\it van Hove singularity} \citep{47} irrespective of the value of spin-wave density, which is decreasing away from half-filling region with growing value of $|\mu|$ at 0.048, one critical point in the flow equation shown in Supplimentary Fig. 2~\cite{SM}). We observe from the phase diagram of $\mu$ and {\it U} at half-filling measured by the prevailing uncertainty from the flow equation (shown in Fig.~\ref{fig_4}). The leading uncertainty region in case of {\it commensurate} spin-wave density is detached from the {\it s}-wave coupling region {\it via} a narrow strip where {\it incommensurate} spin-wave density fluctuations show dominant effect with {\it q}=($\pi$, $\pi$-$\delta$). For small value of {\it U}, the region surrounding half-filling is exponentially small, where uncertainties in spin-wave density play a dominant effect. The value of ({\it t, U}), {\it in-plane} phase diagram is shown in Fig.~\ref{fig_4} with $|\mu|$=4t and $\mu${$<$}0 (below half-filling).\\

\subsection{Behavioral trend of critical energy scale}
\begin{figure}[th!]     
\centering           
\includegraphics[width=8.5cm,height=4.5cm]{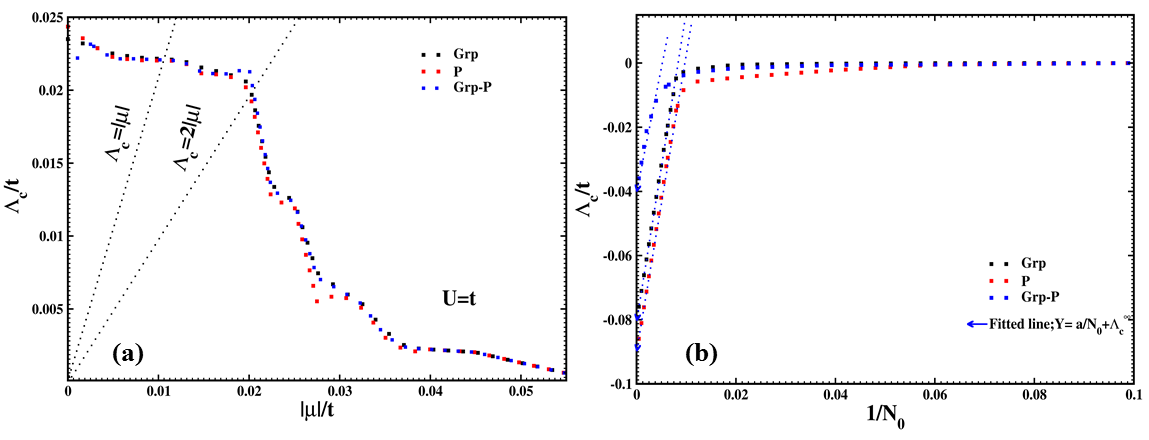}
\caption{The critical energy scale, $\varLambda_c$, as a function of (a) $\mu$ at U=t and n=1. The dotted black lines present the functions for $\varLambda_c$=$|\mu|$ and $\varLambda_c$=2$|\mu|$ for the {\it Grp-P} heterostructure system with a comparison to the pristine graphene and phosphorene sheet. (b) number of discretization points, N$_0$, upon Fermi surface at U=t, n=1 and $|\mu|$=0.005. The dotted black lines represent the functions Y=$\dfrac{a}{(N_0+\varLambda_c)}$.}
 \label{fig_5}
\end{figure}

The behavioral trend of critical energy scale, $\varXi_c$, depending on the function at t$<$0 where, $|\mu|\rightarrow$0 and U=t is plotted in Fig.~\ref{fig_5}(a). The shrinkage in the value of $\varXi_c$ with growing trend of $|\mu|$ confirms that the E$_F$ residues on the {\it van Hove singularity} within the weakly interacting regime having only Fermi surface nesting dominance. Momentum variables of $\Gamma$ have been projected upon the Fermi surfaces at specific points, where the strong renormalization of the vertex function occurs. Here, total 8 fixed points considered for the normalization process. Besides, additional 8 points have been considered in a more reﬁned projection scheme upon the {\it van Hove surface} to check the collective effect.\\

As a result, sum total 16 points (8 for {\it Fermi surface} and 8 for {\it van Hove surface}) have been taken during the flow equation calculations. These refinements, in the response function of the spin-wave density scheme, enhance the analytical calculation efficiency significantly with a reasonable reduction in critical energy scale, without altering $\Gamma$ and $\chi$ values, qualitatively. The variation trend of $\varLambda_c$ as a function of the inverse number of discretization points, N$_0$, upon the Fermi surface at fixed choice of model parameters is shown in Fig.~\ref{fig_5}(b). It is noticed that the critical energy scale is varying in a discrete manner with 8 discretization points, and resumes stability for next 8 points with right order of magnitude indicating the controlled trend of flow equation for weakly interacting systems. This last result shows that the weak coupling behavior of Grp-P system is {\it qualitatively} different from that in pristine graphene. In graphene, the quasiparticle $Y$ factor tends to a finite value at zero energy, \textit{i.e.} the system retains Fermi-liquid behavior with well defined quasiparticles, on the contrary, do not become sharp quasiparticles, even at the lowest energies.\\

\section{Experimental Scope of this study}In this presnt work, we call the Renormalization Group (RG) analysis as the most promising and best controlled analytical approach for low dimensional Fermi systems. In 2D or bulk systems, RG methods are probably the least biased approach to study such systems at weak coupling and low energy scales in a transparent way. This has led to the development of alternative flow scheme. It provides the possibility to tackle the competing instabilities from different angle. It agrees quite favorably over a wider parameter range of models like the 2D Hubbard model. This connection of RG method and 2D Hubbard model encourages to further proceed on a firm approach to understand the instabilities in 2D system. In this present study, we have considered van der Waals heterostructure system, where RG Theory and 2D Hubbard model are implemented for practical realization to explore the concept of spin-wave assisted device application supporting strongly towards the experimental realization. \\

\begin{figure}[th]     
\centering           
\includegraphics[width=8.5cm,height=8.0cm]{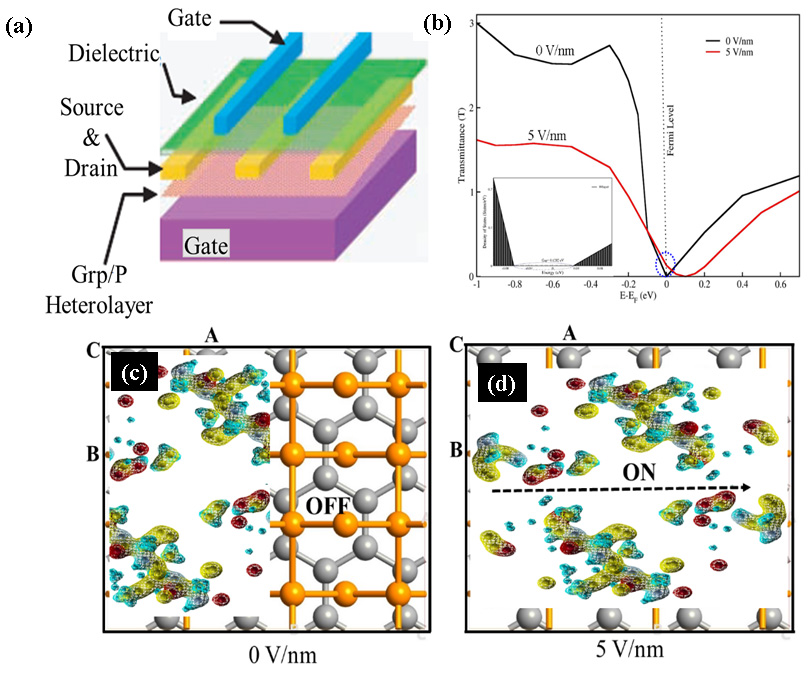}
\caption{(a) Illustration of the chosen dual-gate FET formed from Grp-P hetero bilayer. Two gates are stationary at top and bottom followed by one dielectric layer in between. (b) Transmission bands at 0 V/nm (black line) and 5 V/nm (red line). Inset: the DOS of the hetero bilayer system (in the scattering region) at 0 V/nm at 300 K. Transmission eigenstates at E$_F$ and $\varGamma$(0, 0) point in the reciprocal {\it k} space at (c) {\it off} state (0 V/nm) and (d) {\it on} state (5 V/nm) at room temperature.}
 \label{fig_6}
\end{figure}

To realize the practical application of this system, we demonstrate the hetero bilayer dual gate field effect transistors (DG-FETs) with high {\it on-off} ratio considering Grp-P bilayer as a model system (shown in Fig.~\ref{fig_6}(a)). The mechanism of Giant Stark Effect is implemented here to design the vdW heterolayer-based dual-gate FET, where unsaturated Grp-P bilayers are acting as electrodes~ \cite{PhysRevB.92.035436,PhysRevB.90.085402} at room temperature. In this aspect, direct device application through RG theory is unrealistic in pristine monolayer form, until and unless corroborated with heterolayer system using {\it ab initio} pseudopotential method (refer to \cite{SM}). It is not a secret that there is no experimental evidence in condensed matter physics for the realization of the model solenoid that could produce the field in monolayer system. Besides, the ability to realize such specified device can be proceeded via perturbed system using a pseudopotential method ({\it i.e.} DFT simulation), which is similar to the vector field potential~\cite{Dolde2011}. Here, saturated Grp-P bilayers serve as electrode-cum-tunneling barriers in the crucial part of the model device having dual ({\it i.e.} top and bottom) gates to generate a transverse electric field. Besides, several devices utilize an orbital field to achieve spin splitting~\cite{PhysRevB.97.220404,PhysRevB.97.125423} via the interference of the spinor wave functions in the 2D systems. Moreover, few more reports are there showing implications of this vector field corroborated with {\it ab initio} DFT simulations both experimentally and theoretically for device applications~\cite{Gazibegovic2017,Castillo2015}. It is evident that all phosphorus based FETs can neglect the metal to semiconductor contact at the interface for this transport behaviour. The transmission band spectrum of the model dual-gate FET at 0 V/nm and 5 V/nm, electric fields free from source to drain voltage is shown in Fig.~\ref{fig_6}(b) at 300 K. The semiconducting characteristic of the saturated Grp-P heterolayer blocks the transmission process near the Fermi level at 0 V/nm electric field, originating a finite transmission gap of 1.12 eV, as a result, there is no transmission states near Fermi level. This absence of states indicates the off state of the device. Similarly, a transmission peak emerges near the Fermi level with a wide range of varying dispersion ({\it i.e.} -0.07 to +0.08 eV), at 5 V/nm electric field value, which is an unusual {\it Van Hove-like singularity} in 2D materials. This supports the on state of the designed dual-gate FET, which is stable at 300 K. The transmission eigenchannels at E and at the $\varGamma$(0, 0) point of the reciprocal {\it k}-space, presented in Fig.~\ref{fig_6}(c) $\&$ (d), intensely illustrate the off and the on state of the designed FET controlled by a dual-gate induced electric field, respectively.\\
Similarly, an external transverse electric field of 0 V/nm, the transmission eigenvalue achieves the value of 0.0012G$_0$ , where (G$_0$ = 2e$^2$/h) depends on electron charge {\it e} and Planck’s constant {\it h}. As a result, the transmission channels have been blocked showing an {\it off} state. In contrast, an external transverse field of 5 V/nm, the transmission eigenvalue extents the value upto 1G$_0$
resulting an open state of the transmission channels ({\it i.e.} on state). The ratio of the {\it on-off} state
is found to be 10$^2$ which indicates the persistent of quantum confinement effect due to nanostructuring in the vdW heterostructure. The value of this model hetero bilayer based dual-gate FET is 3 order higher than pristine phosphorene FETs~\cite{Koenig2014} and 1 order higher compared to graphene-MoS$_2$ based FETs at nanoscale~\cite{Koenig2014,Li2014}. The calculated density of states of the scattering region at 5 V/nm as an external electric field is inset in Fig.~\ref{fig_6}(b). The calculated DOS pattern indicates a distinct finite peak near the Fermi level, implying a strong correlation between the transmission channels. The physics behind such strong correlation shows that the transport phenomena is mainly dominated by resonant tunneling through interface states [shown in Fig.~\ref{fig_6}(c) and (d)], not barrier tunneling at the Fermi level.\\

\section{SUMMARY AND CONCLUSIONS}
A study on the spin-wave response function behavior of vdW Grp-P heterostructure was presented via field-theory perspective implementing analytical RG theory. This formulation can be used systematically to identify coupling uncertainties in 2D weakly interacting Fermi systems. In this system, critical energy can be scaled at specific singular points where vertex functions and susceptibilities diverge and become finite. Moreover, the instabilities are modulated via short range antiferromagnetic correlations present in the system, which is noticed from the flow of vertex functions and spontaneous susceptibility. The analytical results show an anomalous singularity of \textit{van Hove} kind in 2D heterostructure. This finding supports the qualitative understanding of vdW heterostructure systems from the anomalous critical points of the response function. This will also facilitate quantitative analysis of room temperature stability of such systems and \textit{on-state} electronic device modeling in future with a step towards experimental realization.

\section*{SUPPLEMENTARY MATERIAL}
See the Supplimentary Material for the physical variables and related equations relating with the phase diagram of the graphene-phosphorene heterostructure system.
\section*{acknowledgments}
SKB acknowledges DST, Govt. of India for providing INSPIRE PhD Fellowship (IF150325). Currently, SKB acknowledges NISER for financial support as a PDF. Part of calculations has been performed in the KALINGA and NISERDFT HPC Facility, NISER. SKB and PD acknowledge Tezpur University for providing HPCC facility. 

\twocolumngrid
\bibliography{rg.bib}

\begin{thebibliography}{75}%
\makeatletter
\providecommand \@ifxundefined [1]{%
 \@ifx{#1\undefined}
}%
\providecommand \@ifnum [1]{%
 \ifnum #1\expandafter \@firstoftwo
 \else \expandafter \@secondoftwo
 \fi
}%
\providecommand \@ifx [1]{%
 \ifx #1\expandafter \@firstoftwo
 \else \expandafter \@secondoftwo
 \fi
}%
\providecommand \natexlab [1]{#1}%
\providecommand \enquote  [1]{``#1''}%
\providecommand \bibnamefont  [1]{#1}%
\providecommand \bibfnamefont [1]{#1}%
\providecommand \citenamefont [1]{#1}%
\providecommand \href@noop [0]{\@secondoftwo}%
\providecommand \href [0]{\begingroup \@sanitize@url \@href}%
\providecommand \@href[1]{\@@startlink{#1}\@@href}%
\providecommand \@@href[1]{\endgroup#1\@@endlink}%
\providecommand \@sanitize@url [0]{\catcode `\\12\catcode `\$12\catcode
  `\&12\catcode `\#12\catcode `\^12\catcode `\_12\catcode `\%12\relax}%
\providecommand \@@startlink[1]{}%
\providecommand \@@endlink[0]{}%
\providecommand \url  [0]{\begingroup\@sanitize@url \@url }%
\providecommand \@url [1]{\endgroup\@href {#1}{\urlprefix }}%
\providecommand \urlprefix  [0]{URL }%
\providecommand \Eprint [0]{\href }%
\providecommand \doibase [0]{http://dx.doi.org/}%
\providecommand \selectlanguage [0]{\@gobble}%
\providecommand \bibinfo  [0]{\@secondoftwo}%
\providecommand \bibfield  [0]{\@secondoftwo}%
\providecommand \translation [1]{[#1]}%
\providecommand \BibitemOpen [0]{}%
\providecommand \bibitemStop [0]{}%
\providecommand \bibitemNoStop [0]{.\EOS\space}%
\providecommand \EOS [0]{\spacefactor3000\relax}%
\providecommand \BibitemShut  [1]{\csname bibitem#1\endcsname}%
\let\auto@bib@innerbib\@empty
\bibitem [{\citenamefont {Nishida}(2016)}]{1}%
  \BibitemOpen
  \bibfield  {author} {\bibinfo {author} {\bibfnamefont {Y.}~\bibnamefont
  {Nishida}},\ }\href {\doibase 10.1103/PhysRevB.94.085430} {\bibfield
  {journal} {\bibinfo  {journal} {Phys. Rev. B}\ }\textbf {\bibinfo {volume}
  {94}},\ \bibinfo {pages} {085430} (\bibinfo {year} {2016})}\BibitemShut
  {NoStop}%
\bibitem [{\citenamefont {Raju}\ \emph {et~al.}(2019)\citenamefont {Raju},
  \citenamefont {Clement}, \citenamefont {Hayden}, \citenamefont {Kent-Dobias},
  \citenamefont {Liarte}, \citenamefont {Rocklin},\ and\ \citenamefont
  {Sethna}}]{2}%
  \BibitemOpen
  \bibfield  {author} {\bibinfo {author} {\bibfnamefont {A.}~\bibnamefont
  {Raju}}, \bibinfo {author} {\bibfnamefont {C.~B.}\ \bibnamefont {Clement}},
  \bibinfo {author} {\bibfnamefont {L.~X.}\ \bibnamefont {Hayden}}, \bibinfo
  {author} {\bibfnamefont {J.~P.}\ \bibnamefont {Kent-Dobias}}, \bibinfo
  {author} {\bibfnamefont {D.~B.}\ \bibnamefont {Liarte}}, \bibinfo {author}
  {\bibfnamefont {D.~Z.}\ \bibnamefont {Rocklin}}, \ and\ \bibinfo {author}
  {\bibfnamefont {J.~P.}\ \bibnamefont {Sethna}},\ }\href {\doibase
  10.1103/PhysRevX.9.021014} {\bibfield  {journal} {\bibinfo  {journal} {Phys.
  Rev. X}\ }\textbf {\bibinfo {volume} {9}},\ \bibinfo {pages} {021014}
  (\bibinfo {year} {2019})}\BibitemShut {NoStop}%
\bibitem [{\citenamefont {Geim}(2009)}]{3}%
  \BibitemOpen
  \bibfield  {author} {\bibinfo {author} {\bibfnamefont {A.~K.}\ \bibnamefont
  {Geim}},\ }\href {\doibase 10.1126/science.1158877} {\bibfield  {journal}
  {\bibinfo  {journal} {Science}\ }\textbf {\bibinfo {volume} {324}},\ \bibinfo
  {pages} {1530} (\bibinfo {year} {2009})}\BibitemShut {NoStop}%
\bibitem [{\citenamefont {Zhang}\ \emph {et~al.}(2005)\citenamefont {Zhang},
  \citenamefont {Small}, \citenamefont {Amori},\ and\ \citenamefont {Kim}}]{5}%
  \BibitemOpen
  \bibfield  {author} {\bibinfo {author} {\bibfnamefont {Y.}~\bibnamefont
  {Zhang}}, \bibinfo {author} {\bibfnamefont {J.~P.}\ \bibnamefont {Small}},
  \bibinfo {author} {\bibfnamefont {M.~E.~S.}\ \bibnamefont {Amori}}, \ and\
  \bibinfo {author} {\bibfnamefont {P.}~\bibnamefont {Kim}},\ }\href
  {https://doi.org/10.1038/nature04233} {\bibfield  {journal} {\bibinfo
  {journal} {Phys. Rev. Lett.}\ }\textbf {\bibinfo {volume} {94}},\ \bibinfo
  {pages} {176803} (\bibinfo {year} {2005})}\BibitemShut {NoStop}%
\bibitem [{\citenamefont {Song}\ \emph {et~al.}(2013)\citenamefont {Song},
  \citenamefont {Shytov},\ and\ \citenamefont
  {Levitov}}]{PhysRevLett.111.266801}%
  \BibitemOpen
  \bibfield  {author} {\bibinfo {author} {\bibfnamefont {J.~C.~W.}\
  \bibnamefont {Song}}, \bibinfo {author} {\bibfnamefont {A.~V.}\ \bibnamefont
  {Shytov}}, \ and\ \bibinfo {author} {\bibfnamefont {L.~S.}\ \bibnamefont
  {Levitov}},\ }\href {\doibase 10.1103/PhysRevLett.111.266801} {\bibfield
  {journal} {\bibinfo  {journal} {Phys. Rev. Lett.}\ }\textbf {\bibinfo
  {volume} {111}},\ \bibinfo {pages} {266801} (\bibinfo {year}
  {2013})}\BibitemShut {NoStop}%
\bibitem [{\citenamefont {Zhang}\ and\ \citenamefont
  {Senthil}(2019)}]{PhysRevB.99.205150}%
  \BibitemOpen
  \bibfield  {author} {\bibinfo {author} {\bibfnamefont {Y.-H.}\ \bibnamefont
  {Zhang}}\ and\ \bibinfo {author} {\bibfnamefont {T.}~\bibnamefont
  {Senthil}},\ }\href {\doibase 10.1103/PhysRevB.99.205150} {\bibfield
  {journal} {\bibinfo  {journal} {Phys. Rev. B}\ }\textbf {\bibinfo {volume}
  {99}},\ \bibinfo {pages} {205150} (\bibinfo {year} {2019})}\BibitemShut
  {NoStop}%
\bibitem [{\citenamefont {Szasz}\ \emph {et~al.}(2020)\citenamefont {Szasz},
  \citenamefont {Motruk}, \citenamefont {Zaletel},\ and\ \citenamefont
  {Moore}}]{PhysRevX.10.021042}%
  \BibitemOpen
  \bibfield  {author} {\bibinfo {author} {\bibfnamefont {A.}~\bibnamefont
  {Szasz}}, \bibinfo {author} {\bibfnamefont {J.}~\bibnamefont {Motruk}},
  \bibinfo {author} {\bibfnamefont {M.~P.}\ \bibnamefont {Zaletel}}, \ and\
  \bibinfo {author} {\bibfnamefont {J.~E.}\ \bibnamefont {Moore}},\ }\href
  {\doibase 10.1103/PhysRevX.10.021042} {\bibfield  {journal} {\bibinfo
  {journal} {Phys. Rev. X}\ }\textbf {\bibinfo {volume} {10}},\ \bibinfo
  {pages} {021042} (\bibinfo {year} {2020})}\BibitemShut {NoStop}%
\bibitem [{\citenamefont {Pan}\ \emph {et~al.}(2020)\citenamefont {Pan},
  \citenamefont {Wu},\ and\ \citenamefont
  {Das~Sarma}}]{PhysRevResearch.2.033087}%
  \BibitemOpen
  \bibfield  {author} {\bibinfo {author} {\bibfnamefont {H.}~\bibnamefont
  {Pan}}, \bibinfo {author} {\bibfnamefont {F.}~\bibnamefont {Wu}}, \ and\
  \bibinfo {author} {\bibfnamefont {S.}~\bibnamefont {Das~Sarma}},\ }\href
  {\doibase 10.1103/PhysRevResearch.2.033087} {\bibfield  {journal} {\bibinfo
  {journal} {Phys. Rev. Research}\ }\textbf {\bibinfo {volume} {2}},\ \bibinfo
  {pages} {033087} (\bibinfo {year} {2020})}\BibitemShut {NoStop}%
\bibitem [{\citenamefont {Wu}\ \emph {et~al.}(2019)\citenamefont {Wu},
  \citenamefont {Lovorn}, \citenamefont {Tutuc}, \citenamefont {Martin},\ and\
  \citenamefont {MacDonald}}]{PhysRevLett.122.086402}%
  \BibitemOpen
  \bibfield  {author} {\bibinfo {author} {\bibfnamefont {F.}~\bibnamefont
  {Wu}}, \bibinfo {author} {\bibfnamefont {T.}~\bibnamefont {Lovorn}}, \bibinfo
  {author} {\bibfnamefont {E.}~\bibnamefont {Tutuc}}, \bibinfo {author}
  {\bibfnamefont {I.}~\bibnamefont {Martin}}, \ and\ \bibinfo {author}
  {\bibfnamefont {A.~H.}\ \bibnamefont {MacDonald}},\ }\href {\doibase
  10.1103/PhysRevLett.122.086402} {\bibfield  {journal} {\bibinfo  {journal}
  {Phys. Rev. Lett.}\ }\textbf {\bibinfo {volume} {122}},\ \bibinfo {pages}
  {086402} (\bibinfo {year} {2019})}\BibitemShut {NoStop}%
\bibitem [{\citenamefont {Wu}\ \emph {et~al.}(2018)\citenamefont {Wu},
  \citenamefont {Lovorn}, \citenamefont {Tutuc},\ and\ \citenamefont
  {MacDonald}}]{PhysRevLett.121.026402}%
  \BibitemOpen
  \bibfield  {author} {\bibinfo {author} {\bibfnamefont {F.}~\bibnamefont
  {Wu}}, \bibinfo {author} {\bibfnamefont {T.}~\bibnamefont {Lovorn}}, \bibinfo
  {author} {\bibfnamefont {E.}~\bibnamefont {Tutuc}}, \ and\ \bibinfo {author}
  {\bibfnamefont {A.~H.}\ \bibnamefont {MacDonald}},\ }\href {\doibase
  10.1103/PhysRevLett.121.026402} {\bibfield  {journal} {\bibinfo  {journal}
  {Phys. Rev. Lett.}\ }\textbf {\bibinfo {volume} {121}},\ \bibinfo {pages}
  {026402} (\bibinfo {year} {2018})}\BibitemShut {NoStop}%
\bibitem [{\citenamefont {Po}\ \emph {et~al.}(2018)\citenamefont {Po},
  \citenamefont {Zou}, \citenamefont {Vishwanath},\ and\ \citenamefont
  {Senthil}}]{PhysRevX.8.031089}%
  \BibitemOpen
  \bibfield  {author} {\bibinfo {author} {\bibfnamefont {H.~C.}\ \bibnamefont
  {Po}}, \bibinfo {author} {\bibfnamefont {L.}~\bibnamefont {Zou}}, \bibinfo
  {author} {\bibfnamefont {A.}~\bibnamefont {Vishwanath}}, \ and\ \bibinfo
  {author} {\bibfnamefont {T.}~\bibnamefont {Senthil}},\ }\href {\doibase
  10.1103/PhysRevX.8.031089} {\bibfield  {journal} {\bibinfo  {journal} {Phys.
  Rev. X}\ }\textbf {\bibinfo {volume} {8}},\ \bibinfo {pages} {031089}
  (\bibinfo {year} {2018})}\BibitemShut {NoStop}%
\bibitem [{\citenamefont {Cao}(2018)}]{Cao2018}%
  \BibitemOpen
  \bibfield  {author} {\bibinfo {author} {\bibfnamefont {Y.~e.~a.}\
  \bibnamefont {Cao}},\ }\href {\doibase 10.1038/nature26154} {\bibfield
  {journal} {\bibinfo  {journal} {Nature}\ }\textbf {\bibinfo {volume} {556}},\
  \bibinfo {pages} {80} (\bibinfo {year} {2018})}\BibitemShut {NoStop}%
\bibitem [{\citenamefont {Salmhofer}(1998{\natexlab{a}})}]{Salmhofer1998}%
  \BibitemOpen
  \bibfield  {author} {\bibinfo {author} {\bibfnamefont {M.}~\bibnamefont
  {Salmhofer}},\ }\href {\doibase 10.1007/s002200050358} {\bibfield  {journal}
  {\bibinfo  {journal} {Communications in Mathematical Physics}\ }\textbf
  {\bibinfo {volume} {194}},\ \bibinfo {pages} {249} (\bibinfo {year}
  {1998}{\natexlab{a}})}\BibitemShut {NoStop}%
\bibitem [{\citenamefont {Gonz\'alez}(2010)}]{PhysRevB.82.155404}%
  \BibitemOpen
  \bibfield  {author} {\bibinfo {author} {\bibfnamefont {J.}~\bibnamefont
  {Gonz\'alez}},\ }\href {\doibase 10.1103/PhysRevB.82.155404} {\bibfield
  {journal} {\bibinfo  {journal} {Phys. Rev. B}\ }\textbf {\bibinfo {volume}
  {82}},\ \bibinfo {pages} {155404} (\bibinfo {year} {2010})}\BibitemShut
  {NoStop}%
\bibitem [{\citenamefont {Katanin}(2013)}]{PhysRevB.88.241401}%
  \BibitemOpen
  \bibfield  {author} {\bibinfo {author} {\bibfnamefont {A.}~\bibnamefont
  {Katanin}},\ }\href {\doibase 10.1103/PhysRevB.88.241401} {\bibfield
  {journal} {\bibinfo  {journal} {Phys. Rev. B}\ }\textbf {\bibinfo {volume}
  {88}},\ \bibinfo {pages} {241401} (\bibinfo {year} {2013})}\BibitemShut
  {NoStop}%
\bibitem [{\citenamefont {Jongeward}\ and\ \citenamefont
  {Wolynes}(1983)}]{Jongeward1983}%
  \BibitemOpen
  \bibfield  {author} {\bibinfo {author} {\bibfnamefont {G.~A.}\ \bibnamefont
  {Jongeward}}\ and\ \bibinfo {author} {\bibfnamefont {P.~G.}\ \bibnamefont
  {Wolynes}},\ }\href {\doibase 10.1063/1.446205} {\bibfield  {journal}
  {\bibinfo  {journal} {The Journal of Chemical Physics}\ }\textbf {\bibinfo
  {volume} {79}},\ \bibinfo {pages} {3517} (\bibinfo {year}
  {1983})}\BibitemShut {NoStop}%
\bibitem [{\citenamefont {Shankar}(1994)}]{30}%
  \BibitemOpen
  \bibfield  {author} {\bibinfo {author} {\bibfnamefont {R.}~\bibnamefont
  {Shankar}},\ }\href {\doibase 10.1103/RevModPhys.66.129} {\bibfield
  {journal} {\bibinfo  {journal} {Rev. Mod. Phys.}\ }\textbf {\bibinfo {volume}
  {66}},\ \bibinfo {pages} {129} (\bibinfo {year} {1994})}\BibitemShut
  {NoStop}%
\bibitem [{\citenamefont {Zanchi}\ and\ \citenamefont
  {Schulz}(2000)}]{PhysRevB.61.13609}%
  \BibitemOpen
  \bibfield  {author} {\bibinfo {author} {\bibfnamefont {D.}~\bibnamefont
  {Zanchi}}\ and\ \bibinfo {author} {\bibfnamefont {H.~J.}\ \bibnamefont
  {Schulz}},\ }\href {\doibase 10.1103/PhysRevB.61.13609} {\bibfield  {journal}
  {\bibinfo  {journal} {Phys. Rev. B}\ }\textbf {\bibinfo {volume} {61}},\
  \bibinfo {pages} {13609} (\bibinfo {year} {2000})}\BibitemShut {NoStop}%
\bibitem [{\citenamefont {Halboth}\ and\ \citenamefont
  {Metzner}(2000{\natexlab{a}})}]{PhysRevB.61.7364}%
  \BibitemOpen
  \bibfield  {author} {\bibinfo {author} {\bibfnamefont {C.~J.}\ \bibnamefont
  {Halboth}}\ and\ \bibinfo {author} {\bibfnamefont {W.}~\bibnamefont
  {Metzner}},\ }\href {\doibase 10.1103/PhysRevB.61.7364} {\bibfield  {journal}
  {\bibinfo  {journal} {Phys. Rev. B}\ }\textbf {\bibinfo {volume} {61}},\
  \bibinfo {pages} {7364} (\bibinfo {year} {2000}{\natexlab{a}})}\BibitemShut
  {NoStop}%
\bibitem [{\citenamefont {Halboth}\ and\ \citenamefont
  {Metzner}(2000{\natexlab{b}})}]{PhysRevLett.85.5162}%
  \BibitemOpen
  \bibfield  {author} {\bibinfo {author} {\bibfnamefont {C.~J.}\ \bibnamefont
  {Halboth}}\ and\ \bibinfo {author} {\bibfnamefont {W.}~\bibnamefont
  {Metzner}},\ }\href {\doibase 10.1103/PhysRevLett.85.5162} {\bibfield
  {journal} {\bibinfo  {journal} {Phys. Rev. Lett.}\ }\textbf {\bibinfo
  {volume} {85}},\ \bibinfo {pages} {5162} (\bibinfo {year}
  {2000}{\natexlab{b}})}\BibitemShut {NoStop}%
\bibitem [{\citenamefont {Honerkamp}\ and\ \citenamefont
  {Salmhofer}(2001)}]{PhysRevLett.87.187004}%
  \BibitemOpen
  \bibfield  {author} {\bibinfo {author} {\bibfnamefont {C.}~\bibnamefont
  {Honerkamp}}\ and\ \bibinfo {author} {\bibfnamefont {M.}~\bibnamefont
  {Salmhofer}},\ }\href {\doibase 10.1103/PhysRevLett.87.187004} {\bibfield
  {journal} {\bibinfo  {journal} {Phys. Rev. Lett.}\ }\textbf {\bibinfo
  {volume} {87}},\ \bibinfo {pages} {187004} (\bibinfo {year}
  {2001})}\BibitemShut {NoStop}%
\bibitem [{\citenamefont {Hsu}\ \emph {et~al.}(2020)\citenamefont {Hsu},
  \citenamefont {Wu},\ and\ \citenamefont {Das~Sarma}}]{PhysRevB.102.085103}%
  \BibitemOpen
  \bibfield  {author} {\bibinfo {author} {\bibfnamefont {Y.-T.}\ \bibnamefont
  {Hsu}}, \bibinfo {author} {\bibfnamefont {F.}~\bibnamefont {Wu}}, \ and\
  \bibinfo {author} {\bibfnamefont {S.}~\bibnamefont {Das~Sarma}},\ }\href
  {\doibase 10.1103/PhysRevB.102.085103} {\bibfield  {journal} {\bibinfo
  {journal} {Phys. Rev. B}\ }\textbf {\bibinfo {volume} {102}},\ \bibinfo
  {pages} {085103} (\bibinfo {year} {2020})}\BibitemShut {NoStop}%
\bibitem [{\citenamefont {Behera}\ and\ \citenamefont
  {Deb}(2020)}]{D0CP00836B}%
  \BibitemOpen
  \bibfield  {author} {\bibinfo {author} {\bibfnamefont {S.~K.}\ \bibnamefont
  {Behera}}\ and\ \bibinfo {author} {\bibfnamefont {P.}~\bibnamefont {Deb}},\
  }\href {\doibase 10.1039/D0CP00836B} {\bibfield  {journal} {\bibinfo
  {journal} {Phys. Chem. Chem. Phys.}\ }\textbf {\bibinfo {volume} {22}},\
  \bibinfo {pages} {19139} (\bibinfo {year} {2020})}\BibitemShut {NoStop}%
\bibitem [{\citenamefont {Hadad}\ and\ \citenamefont {Steinberg}(2010)}]{26}%
  \BibitemOpen
  \bibfield  {author} {\bibinfo {author} {\bibfnamefont {Y.}~\bibnamefont
  {Hadad}}\ and\ \bibinfo {author} {\bibfnamefont {B.~Z.}\ \bibnamefont
  {Steinberg}},\ }\href {\doibase 10.1103/PhysRevLett.105.233904} {\bibfield
  {journal} {\bibinfo  {journal} {Phys. Rev. Lett.}\ }\textbf {\bibinfo
  {volume} {105}},\ \bibinfo {pages} {233904} (\bibinfo {year}
  {2010})}\BibitemShut {NoStop}%
\bibitem [{\citenamefont {Isobe}\ \emph {et~al.}(2016)\citenamefont {Isobe},
  \citenamefont {Yang}, \citenamefont {Chubukov}, \citenamefont {Schmalian},\
  and\ \citenamefont {Nagaosa}}]{PhysRevLett.116.076803}%
  \BibitemOpen
  \bibfield  {author} {\bibinfo {author} {\bibfnamefont {H.}~\bibnamefont
  {Isobe}}, \bibinfo {author} {\bibfnamefont {B.-J.}\ \bibnamefont {Yang}},
  \bibinfo {author} {\bibfnamefont {A.}~\bibnamefont {Chubukov}}, \bibinfo
  {author} {\bibfnamefont {J.}~\bibnamefont {Schmalian}}, \ and\ \bibinfo
  {author} {\bibfnamefont {N.}~\bibnamefont {Nagaosa}},\ }\href {\doibase
  10.1103/PhysRevLett.116.076803} {\bibfield  {journal} {\bibinfo  {journal}
  {Phys. Rev. Lett.}\ }\textbf {\bibinfo {volume} {116}},\ \bibinfo {pages}
  {076803} (\bibinfo {year} {2016})}\BibitemShut {NoStop}%
\bibitem [{\citenamefont {Kozikov}\ \emph {et~al.}(2010)\citenamefont
  {Kozikov}, \citenamefont {Savchenko}, \citenamefont {Narozhny},\ and\
  \citenamefont {Shytov}}]{PhysRevB.82.075424}%
  \BibitemOpen
  \bibfield  {author} {\bibinfo {author} {\bibfnamefont {A.~A.}\ \bibnamefont
  {Kozikov}}, \bibinfo {author} {\bibfnamefont {A.~K.}\ \bibnamefont
  {Savchenko}}, \bibinfo {author} {\bibfnamefont {B.~N.}\ \bibnamefont
  {Narozhny}}, \ and\ \bibinfo {author} {\bibfnamefont {A.~V.}\ \bibnamefont
  {Shytov}},\ }\href {\doibase 10.1103/PhysRevB.82.075424} {\bibfield
  {journal} {\bibinfo  {journal} {Phys. Rev. B}\ }\textbf {\bibinfo {volume}
  {82}},\ \bibinfo {pages} {075424} (\bibinfo {year} {2010})}\BibitemShut
  {NoStop}%
\bibitem [{\citenamefont {Gusynin}\ and\ \citenamefont {Sharapov}(2005)}]{27}%
  \BibitemOpen
  \bibfield  {author} {\bibinfo {author} {\bibfnamefont {V.~P.}\ \bibnamefont
  {Gusynin}}\ and\ \bibinfo {author} {\bibfnamefont {S.~G.}\ \bibnamefont
  {Sharapov}},\ }\href {\doibase 10.1103/PhysRevLett.95.146801} {\bibfield
  {journal} {\bibinfo  {journal} {Phys. Rev. Lett.}\ }\textbf {\bibinfo
  {volume} {95}},\ \bibinfo {pages} {146801} (\bibinfo {year}
  {2005})}\BibitemShut {NoStop}%
\bibitem [{\citenamefont {Wilson}(1975)}]{29}%
  \BibitemOpen
  \bibfield  {author} {\bibinfo {author} {\bibfnamefont {K.~G.}\ \bibnamefont
  {Wilson}},\ }\href {\doibase 10.1103/RevModPhys.47.773} {\bibfield  {journal}
  {\bibinfo  {journal} {Rev. Mod. Phys.}\ }\textbf {\bibinfo {volume} {47}},\
  \bibinfo {pages} {773} (\bibinfo {year} {1975})}\BibitemShut {NoStop}%
\bibitem [{\citenamefont {Naji}\ \emph {et~al.}(2013)\citenamefont {Naji},
  \citenamefont {Kanduč}, \citenamefont {Forsman},\ and\ \citenamefont
  {Podgornik}}]{Naji2013}%
  \BibitemOpen
  \bibfield  {author} {\bibinfo {author} {\bibfnamefont {A.}~\bibnamefont
  {Naji}}, \bibinfo {author} {\bibfnamefont {M.}~\bibnamefont {Kanduč}},
  \bibinfo {author} {\bibfnamefont {J.}~\bibnamefont {Forsman}}, \ and\
  \bibinfo {author} {\bibfnamefont {R.}~\bibnamefont {Podgornik}},\ }\href
  {\doibase 10.1063/1.4824681} {\bibfield  {journal} {\bibinfo  {journal} {The
  Journal of Chemical Physics}\ }\textbf {\bibinfo {volume} {139}},\ \bibinfo
  {pages} {150901} (\bibinfo {year} {2013})}\BibitemShut {NoStop}%
\bibitem [{\citenamefont {Ziletti}\ \emph {et~al.}(2015)\citenamefont
  {Ziletti}, \citenamefont {Huang}, \citenamefont {Coker},\ and\ \citenamefont
  {Lin}}]{PhysRevB.92.085423}%
  \BibitemOpen
  \bibfield  {author} {\bibinfo {author} {\bibfnamefont {A.}~\bibnamefont
  {Ziletti}}, \bibinfo {author} {\bibfnamefont {S.~M.}\ \bibnamefont {Huang}},
  \bibinfo {author} {\bibfnamefont {D.~F.}\ \bibnamefont {Coker}}, \ and\
  \bibinfo {author} {\bibfnamefont {H.}~\bibnamefont {Lin}},\ }\href {\doibase
  10.1103/PhysRevB.92.085423} {\bibfield  {journal} {\bibinfo  {journal} {Phys.
  Rev. B}\ }\textbf {\bibinfo {volume} {92}},\ \bibinfo {pages} {085423}
  (\bibinfo {year} {2015})}\BibitemShut {NoStop}%
\bibitem [{\citenamefont {Bickers}\ \emph
  {et~al.}(1989{\natexlab{a}})\citenamefont {Bickers}, \citenamefont
  {Scalapino},\ and\ \citenamefont {White}}]{31}%
  \BibitemOpen
  \bibfield  {author} {\bibinfo {author} {\bibfnamefont {N.~E.}\ \bibnamefont
  {Bickers}}, \bibinfo {author} {\bibfnamefont {D.~J.}\ \bibnamefont
  {Scalapino}}, \ and\ \bibinfo {author} {\bibfnamefont {S.~R.}\ \bibnamefont
  {White}},\ }\href {\doibase 10.1103/PhysRevLett.62.961} {\bibfield  {journal}
  {\bibinfo  {journal} {Phys. Rev. Lett.}\ }\textbf {\bibinfo {volume} {62}},\
  \bibinfo {pages} {961} (\bibinfo {year} {1989}{\natexlab{a}})}\BibitemShut
  {NoStop}%
\bibitem [{\citenamefont {Bickers}\ \emph
  {et~al.}(1989{\natexlab{b}})\citenamefont {Bickers}, \citenamefont
  {Scalapino},\ and\ \citenamefont {White}}]{32}%
  \BibitemOpen
  \bibfield  {author} {\bibinfo {author} {\bibfnamefont {N.~E.}\ \bibnamefont
  {Bickers}}, \bibinfo {author} {\bibfnamefont {D.~J.}\ \bibnamefont
  {Scalapino}}, \ and\ \bibinfo {author} {\bibfnamefont {S.~R.}\ \bibnamefont
  {White}},\ }\href {\doibase 10.1103/PhysRevLett.62.961} {\bibfield  {journal}
  {\bibinfo  {journal} {Phys. Rev. Lett.}\ }\textbf {\bibinfo {volume} {62}},\
  \bibinfo {pages} {961} (\bibinfo {year} {1989}{\natexlab{b}})}\BibitemShut
  {NoStop}%
\bibitem [{\citenamefont {Furukawa}\ \emph {et~al.}(1998)\citenamefont
  {Furukawa}, \citenamefont {Rice},\ and\ \citenamefont {Salmhofer}}]{35}%
  \BibitemOpen
  \bibfield  {author} {\bibinfo {author} {\bibfnamefont {N.}~\bibnamefont
  {Furukawa}}, \bibinfo {author} {\bibfnamefont {T.~M.}\ \bibnamefont {Rice}},
  \ and\ \bibinfo {author} {\bibfnamefont {M.}~\bibnamefont {Salmhofer}},\
  }\href {\doibase 10.1103/PhysRevLett.81.3195} {\bibfield  {journal} {\bibinfo
   {journal} {Phys. Rev. Lett.}\ }\textbf {\bibinfo {volume} {81}},\ \bibinfo
  {pages} {3195} (\bibinfo {year} {1998})}\BibitemShut {NoStop}%
\bibitem [{\citenamefont {Salmhofer}(1998{\natexlab{b}})}]{36}%
  \BibitemOpen
  \bibfield  {author} {\bibinfo {author} {\bibfnamefont {M.}~\bibnamefont
  {Salmhofer}},\ }\href {\doibase 10.1007/s002200050358} {\bibfield  {journal}
  {\bibinfo  {journal} {Communications in Mathematical Physics}\ }\textbf
  {\bibinfo {volume} {194}},\ \bibinfo {pages} {249} (\bibinfo {year}
  {1998}{\natexlab{b}})}\BibitemShut {NoStop}%
\bibitem [{\citenamefont {Behera}\ \emph {et~al.}(2019)\citenamefont {Behera},
  \citenamefont {Bora}, \citenamefont {Paul~Chowdhury},\ and\ \citenamefont
  {Deb}}]{C9CP05252F}%
  \BibitemOpen
  \bibfield  {author} {\bibinfo {author} {\bibfnamefont {S.~K.}\ \bibnamefont
  {Behera}}, \bibinfo {author} {\bibfnamefont {M.}~\bibnamefont {Bora}},
  \bibinfo {author} {\bibfnamefont {S.~S.}\ \bibnamefont {Paul~Chowdhury}}, \
  and\ \bibinfo {author} {\bibfnamefont {P.}~\bibnamefont {Deb}},\ }\href
  {\doibase 10.1039/C9CP05252F} {\bibfield  {journal} {\bibinfo  {journal}
  {Phys. Chem. Chem. Phys.}\ }\textbf {\bibinfo {volume} {21}},\ \bibinfo
  {pages} {25788} (\bibinfo {year} {2019})}\BibitemShut {NoStop}%
\bibitem [{\citenamefont {{Dalton}}\ \emph {et~al.}(2016)\citenamefont
  {{Dalton}}, \citenamefont {{Jeffers}},\ and\ \citenamefont {{Barnett}}}]{40}%
  \BibitemOpen
  \bibfield  {author} {\bibinfo {author} {\bibfnamefont {B.~J.}\ \bibnamefont
  {{Dalton}}}, \bibinfo {author} {\bibfnamefont {J.}~\bibnamefont {{Jeffers}}},
  \ and\ \bibinfo {author} {\bibfnamefont {S.~M.}\ \bibnamefont {{Barnett}}},\
  }\href {\doibase 10.1016/j.aop.2016.03.006} {\bibfield  {journal} {\bibinfo
  {journal} {Annals of Physics}\ }\textbf {\bibinfo {volume} {370}},\ \bibinfo
  {pages} {12} (\bibinfo {year} {2016})},\ \Eprint
  {http://arxiv.org/abs/1604.03375} {1604.03375} \BibitemShut {NoStop}%
\bibitem [{\citenamefont {describes the~physical variables}\ and\ \citenamefont
  {related equations relating with the phase diagram of the
  graphene-phosphorene heterostructure system. Details of the DFT
  calculations~are explined}()}]{SM}%
  \BibitemOpen
  \bibfield  {author} {\bibinfo {author} {\bibfnamefont {S.~M.}\ \bibnamefont
  {describes the~physical variables}}\ and\ \bibinfo {author} {\bibnamefont
  {related equations relating with the phase diagram of the
  graphene-phosphorene heterostructure system. Details of the DFT
  calculations~are explined}},\ }\href@noop {} {\ }\BibitemShut {NoStop}%
\bibitem [{\citenamefont {Castro~Neto}\ \emph
  {et~al.}(2009{\natexlab{a}})\citenamefont {Castro~Neto}, \citenamefont
  {Guinea}, \citenamefont {Peres}, \citenamefont {Novoselov},\ and\
  \citenamefont {Geim}}]{RevModPhys.81.109}%
  \BibitemOpen
  \bibfield  {author} {\bibinfo {author} {\bibfnamefont {A.~H.}\ \bibnamefont
  {Castro~Neto}}, \bibinfo {author} {\bibfnamefont {F.}~\bibnamefont {Guinea}},
  \bibinfo {author} {\bibfnamefont {N.~M.~R.}\ \bibnamefont {Peres}}, \bibinfo
  {author} {\bibfnamefont {K.~S.}\ \bibnamefont {Novoselov}}, \ and\ \bibinfo
  {author} {\bibfnamefont {A.~K.}\ \bibnamefont {Geim}},\ }\href {\doibase
  10.1103/RevModPhys.81.109} {\bibfield  {journal} {\bibinfo  {journal} {Rev.
  Mod. Phys.}\ }\textbf {\bibinfo {volume} {81}},\ \bibinfo {pages} {109}
  (\bibinfo {year} {2009}{\natexlab{a}})}\BibitemShut {NoStop}%
\bibitem [{\citenamefont {Basko}\ and\ \citenamefont
  {Aleiner}(2008)}]{PhysRevB.77.041409}%
  \BibitemOpen
  \bibfield  {author} {\bibinfo {author} {\bibfnamefont {D.~M.}\ \bibnamefont
  {Basko}}\ and\ \bibinfo {author} {\bibfnamefont {I.~L.}\ \bibnamefont
  {Aleiner}},\ }\href {\doibase 10.1103/PhysRevB.77.041409} {\bibfield
  {journal} {\bibinfo  {journal} {Phys. Rev. B}\ }\textbf {\bibinfo {volume}
  {77}},\ \bibinfo {pages} {041409} (\bibinfo {year} {2008})}\BibitemShut
  {NoStop}%
\bibitem [{\citenamefont {Foster}\ and\ \citenamefont
  {Aleiner}(2008)}]{PhysRevB.77.195413}%
  \BibitemOpen
  \bibfield  {author} {\bibinfo {author} {\bibfnamefont {M.~S.}\ \bibnamefont
  {Foster}}\ and\ \bibinfo {author} {\bibfnamefont {I.~L.}\ \bibnamefont
  {Aleiner}},\ }\href {\doibase 10.1103/PhysRevB.77.195413} {\bibfield
  {journal} {\bibinfo  {journal} {Phys. Rev. B}\ }\textbf {\bibinfo {volume}
  {77}},\ \bibinfo {pages} {195413} (\bibinfo {year} {2008})}\BibitemShut
  {NoStop}%
\bibitem [{\citenamefont {Murray}\ and\ \citenamefont
  {Vafek}(2014)}]{PhysRevB.89.201110}%
  \BibitemOpen
  \bibfield  {author} {\bibinfo {author} {\bibfnamefont {J.~M.}\ \bibnamefont
  {Murray}}\ and\ \bibinfo {author} {\bibfnamefont {O.}~\bibnamefont {Vafek}},\
  }\href {\doibase 10.1103/PhysRevB.89.201110} {\bibfield  {journal} {\bibinfo
  {journal} {Phys. Rev. B}\ }\textbf {\bibinfo {volume} {89}},\ \bibinfo
  {pages} {201110} (\bibinfo {year} {2014})}\BibitemShut {NoStop}%
\bibitem [{\citenamefont {Sheehy}\ and\ \citenamefont
  {Schmalian}(2007)}]{PhysRevLett.99.226803}%
  \BibitemOpen
  \bibfield  {author} {\bibinfo {author} {\bibfnamefont {D.~E.}\ \bibnamefont
  {Sheehy}}\ and\ \bibinfo {author} {\bibfnamefont {J.}~\bibnamefont
  {Schmalian}},\ }\href {\doibase 10.1103/PhysRevLett.99.226803} {\bibfield
  {journal} {\bibinfo  {journal} {Phys. Rev. Lett.}\ }\textbf {\bibinfo
  {volume} {99}},\ \bibinfo {pages} {226803} (\bibinfo {year}
  {2007})}\BibitemShut {NoStop}%
\bibitem [{\citenamefont {Goswami}\ and\ \citenamefont
  {Chakravarty}(2011)}]{PhysRevLett.107.196803}%
  \BibitemOpen
  \bibfield  {author} {\bibinfo {author} {\bibfnamefont {P.}~\bibnamefont
  {Goswami}}\ and\ \bibinfo {author} {\bibfnamefont {S.}~\bibnamefont
  {Chakravarty}},\ }\href {\doibase 10.1103/PhysRevLett.107.196803} {\bibfield
  {journal} {\bibinfo  {journal} {Phys. Rev. Lett.}\ }\textbf {\bibinfo
  {volume} {107}},\ \bibinfo {pages} {196803} (\bibinfo {year}
  {2011})}\BibitemShut {NoStop}%
\bibitem [{\citenamefont {Hosur}\ \emph {et~al.}(2012)\citenamefont {Hosur},
  \citenamefont {Parameswaran},\ and\ \citenamefont
  {Vishwanath}}]{PhysRevLett.108.046602}%
  \BibitemOpen
  \bibfield  {author} {\bibinfo {author} {\bibfnamefont {P.}~\bibnamefont
  {Hosur}}, \bibinfo {author} {\bibfnamefont {S.~A.}\ \bibnamefont
  {Parameswaran}}, \ and\ \bibinfo {author} {\bibfnamefont {A.}~\bibnamefont
  {Vishwanath}},\ }\href {\doibase 10.1103/PhysRevLett.108.046602} {\bibfield
  {journal} {\bibinfo  {journal} {Phys. Rev. Lett.}\ }\textbf {\bibinfo
  {volume} {108}},\ \bibinfo {pages} {046602} (\bibinfo {year}
  {2012})}\BibitemShut {NoStop}%
\bibitem [{\citenamefont {Isobe}\ and\ \citenamefont
  {Nagaosa}(2012)}]{PhysRevB.86.165127}%
  \BibitemOpen
  \bibfield  {author} {\bibinfo {author} {\bibfnamefont {H.}~\bibnamefont
  {Isobe}}\ and\ \bibinfo {author} {\bibfnamefont {N.}~\bibnamefont
  {Nagaosa}},\ }\href {\doibase 10.1103/PhysRevB.86.165127} {\bibfield
  {journal} {\bibinfo  {journal} {Phys. Rev. B}\ }\textbf {\bibinfo {volume}
  {86}},\ \bibinfo {pages} {165127} (\bibinfo {year} {2012})}\BibitemShut
  {NoStop}%
\bibitem [{\citenamefont {Isobe}\ and\ \citenamefont
  {Nagaosa}(2013)}]{PhysRevB.87.205138}%
  \BibitemOpen
  \bibfield  {author} {\bibinfo {author} {\bibfnamefont {H.}~\bibnamefont
  {Isobe}}\ and\ \bibinfo {author} {\bibfnamefont {N.}~\bibnamefont
  {Nagaosa}},\ }\href {\doibase 10.1103/PhysRevB.87.205138} {\bibfield
  {journal} {\bibinfo  {journal} {Phys. Rev. B}\ }\textbf {\bibinfo {volume}
  {87}},\ \bibinfo {pages} {205138} (\bibinfo {year} {2013})}\BibitemShut
  {NoStop}%
\bibitem [{\citenamefont {Sheehy}\ and\ \citenamefont
  {Schmalian}(2009)}]{PhysRevB.80.193411}%
  \BibitemOpen
  \bibfield  {author} {\bibinfo {author} {\bibfnamefont {D.~E.}\ \bibnamefont
  {Sheehy}}\ and\ \bibinfo {author} {\bibfnamefont {J.}~\bibnamefont
  {Schmalian}},\ }\href {\doibase 10.1103/PhysRevB.80.193411} {\bibfield
  {journal} {\bibinfo  {journal} {Phys. Rev. B}\ }\textbf {\bibinfo {volume}
  {80}},\ \bibinfo {pages} {193411} (\bibinfo {year} {2009})}\BibitemShut
  {NoStop}%
\bibitem [{\citenamefont {Hur}\ and\ \citenamefont {{Maurice
  Rice}}(2009)}]{HUR20091452}%
  \BibitemOpen
  \bibfield  {author} {\bibinfo {author} {\bibfnamefont {K.~L.}\ \bibnamefont
  {Hur}}\ and\ \bibinfo {author} {\bibfnamefont {T.}~\bibnamefont {{Maurice
  Rice}}},\ }\href {\doibase https://doi.org/10.1016/j.aop.2009.02.004}
  {\bibfield  {journal} {\bibinfo  {journal} {Annals of Physics}\ }\textbf
  {\bibinfo {volume} {324}},\ \bibinfo {pages} {1452 } (\bibinfo {year}
  {2009})},\ \bibinfo {note} {july 2009 Special Issue}\BibitemShut {NoStop}%
\bibitem [{\citenamefont {Metlitski}\ and\ \citenamefont {Sachdev}(2010)}]{42}%
  \BibitemOpen
  \bibfield  {author} {\bibinfo {author} {\bibfnamefont {M.~A.}\ \bibnamefont
  {Metlitski}}\ and\ \bibinfo {author} {\bibfnamefont {S.}~\bibnamefont
  {Sachdev}},\ }\href {\doibase 10.1103/PhysRevB.82.075128} {\bibfield
  {journal} {\bibinfo  {journal} {Phys. Rev. B}\ }\textbf {\bibinfo {volume}
  {82}},\ \bibinfo {pages} {075128} (\bibinfo {year} {2010})}\BibitemShut
  {NoStop}%
\bibitem [{\citenamefont {Chitov}\ and\ \citenamefont
  {S\'en\'echal}(1998)}]{47}%
  \BibitemOpen
  \bibfield  {author} {\bibinfo {author} {\bibfnamefont {G.~Y.}\ \bibnamefont
  {Chitov}}\ and\ \bibinfo {author} {\bibfnamefont {D.}~\bibnamefont
  {S\'en\'echal}},\ }\href {\doibase 10.1103/PhysRevB.57.1444} {\bibfield
  {journal} {\bibinfo  {journal} {Phys. Rev. B}\ }\textbf {\bibinfo {volume}
  {57}},\ \bibinfo {pages} {1444} (\bibinfo {year} {1998})}\BibitemShut
  {NoStop}%
\bibitem [{\citenamefont {Wu}\ \emph {et~al.}(2015)\citenamefont {Wu},
  \citenamefont {Shen}, \citenamefont {Yang}, \citenamefont {Cai},
  \citenamefont {Huang},\ and\ \citenamefont {Feng}}]{PhysRevB.92.035436}%
  \BibitemOpen
  \bibfield  {author} {\bibinfo {author} {\bibfnamefont {Q.}~\bibnamefont
  {Wu}}, \bibinfo {author} {\bibfnamefont {L.}~\bibnamefont {Shen}}, \bibinfo
  {author} {\bibfnamefont {M.}~\bibnamefont {Yang}}, \bibinfo {author}
  {\bibfnamefont {Y.}~\bibnamefont {Cai}}, \bibinfo {author} {\bibfnamefont
  {Z.}~\bibnamefont {Huang}}, \ and\ \bibinfo {author} {\bibfnamefont {Y.~P.}\
  \bibnamefont {Feng}},\ }\href {\doibase 10.1103/PhysRevB.92.035436}
  {\bibfield  {journal} {\bibinfo  {journal} {Phys. Rev. B}\ }\textbf {\bibinfo
  {volume} {92}},\ \bibinfo {pages} {035436} (\bibinfo {year}
  {2015})}\BibitemShut {NoStop}%
\bibitem [{\citenamefont {Peng}\ \emph {et~al.}(2014)\citenamefont {Peng},
  \citenamefont {Wei},\ and\ \citenamefont {Copple}}]{PhysRevB.90.085402}%
  \BibitemOpen
  \bibfield  {author} {\bibinfo {author} {\bibfnamefont {X.}~\bibnamefont
  {Peng}}, \bibinfo {author} {\bibfnamefont {Q.}~\bibnamefont {Wei}}, \ and\
  \bibinfo {author} {\bibfnamefont {A.}~\bibnamefont {Copple}},\ }\href
  {\doibase 10.1103/PhysRevB.90.085402} {\bibfield  {journal} {\bibinfo
  {journal} {Phys. Rev. B}\ }\textbf {\bibinfo {volume} {90}},\ \bibinfo
  {pages} {085402} (\bibinfo {year} {2014})}\BibitemShut {NoStop}%
\bibitem [{\citenamefont {Dolde}\ \emph {et~al.}(2011)\citenamefont {Dolde},
  \citenamefont {Fedder}, \citenamefont {Doherty}, \citenamefont {N{\"o}bauer},
  \citenamefont {Rempp}, \citenamefont {Balasubramanian}, \citenamefont {Wolf},
  \citenamefont {Reinhard}, \citenamefont {Hollenberg}, \citenamefont
  {Jelezko},\ and\ \citenamefont {Wrachtrup}}]{Dolde2011}%
  \BibitemOpen
  \bibfield  {author} {\bibinfo {author} {\bibfnamefont {F.}~\bibnamefont
  {Dolde}}, \bibinfo {author} {\bibfnamefont {H.}~\bibnamefont {Fedder}},
  \bibinfo {author} {\bibfnamefont {M.~W.}\ \bibnamefont {Doherty}}, \bibinfo
  {author} {\bibfnamefont {T.}~\bibnamefont {N{\"o}bauer}}, \bibinfo {author}
  {\bibfnamefont {F.}~\bibnamefont {Rempp}}, \bibinfo {author} {\bibfnamefont
  {G.}~\bibnamefont {Balasubramanian}}, \bibinfo {author} {\bibfnamefont
  {T.}~\bibnamefont {Wolf}}, \bibinfo {author} {\bibfnamefont {F.}~\bibnamefont
  {Reinhard}}, \bibinfo {author} {\bibfnamefont {L.~C.~L.}\ \bibnamefont
  {Hollenberg}}, \bibinfo {author} {\bibfnamefont {F.}~\bibnamefont {Jelezko}},
  \ and\ \bibinfo {author} {\bibfnamefont {J.}~\bibnamefont {Wrachtrup}},\
  }\href {\doibase 10.1038/nphys1969} {\bibfield  {journal} {\bibinfo
  {journal} {Nature Physics}\ }\textbf {\bibinfo {volume} {7}},\ \bibinfo
  {pages} {459} (\bibinfo {year} {2011})}\BibitemShut {NoStop}%
\bibitem [{\citenamefont {Aharony}\ \emph {et~al.}(2018)\citenamefont
  {Aharony}, \citenamefont {Entin-Wohlman}, \citenamefont {Jonson},\ and\
  \citenamefont {Shekhter}}]{PhysRevB.97.220404}%
  \BibitemOpen
  \bibfield  {author} {\bibinfo {author} {\bibfnamefont {A.}~\bibnamefont
  {Aharony}}, \bibinfo {author} {\bibfnamefont {O.}~\bibnamefont
  {Entin-Wohlman}}, \bibinfo {author} {\bibfnamefont {M.}~\bibnamefont
  {Jonson}}, \ and\ \bibinfo {author} {\bibfnamefont {R.~I.}\ \bibnamefont
  {Shekhter}},\ }\href {\doibase 10.1103/PhysRevB.97.220404} {\bibfield
  {journal} {\bibinfo  {journal} {Phys. Rev. B}\ }\textbf {\bibinfo {volume}
  {97}},\ \bibinfo {pages} {220404} (\bibinfo {year} {2018})}\BibitemShut
  {NoStop}%
\bibitem [{\citenamefont {Saarikoski}\ \emph {et~al.}(2018)\citenamefont
  {Saarikoski}, \citenamefont {Reynoso}, \citenamefont {Baltan\'as},
  \citenamefont {Frustaglia},\ and\ \citenamefont
  {Nitta}}]{PhysRevB.97.125423}%
  \BibitemOpen
  \bibfield  {author} {\bibinfo {author} {\bibfnamefont {H.}~\bibnamefont
  {Saarikoski}}, \bibinfo {author} {\bibfnamefont {A.~A.}\ \bibnamefont
  {Reynoso}}, \bibinfo {author} {\bibfnamefont {J.~P.}\ \bibnamefont
  {Baltan\'as}}, \bibinfo {author} {\bibfnamefont {D.}~\bibnamefont
  {Frustaglia}}, \ and\ \bibinfo {author} {\bibfnamefont {J.}~\bibnamefont
  {Nitta}},\ }\href {\doibase 10.1103/PhysRevB.97.125423} {\bibfield  {journal}
  {\bibinfo  {journal} {Phys. Rev. B}\ }\textbf {\bibinfo {volume} {97}},\
  \bibinfo {pages} {125423} (\bibinfo {year} {2018})}\BibitemShut {NoStop}%
\bibitem [{\citenamefont {Gazibegovic}\ \emph {et~al.}(2017)\citenamefont
  {Gazibegovic}, \citenamefont {Car}, \citenamefont {Zhang}, \citenamefont
  {Balk}, \citenamefont {Logan}, \citenamefont {de~Moor}, \citenamefont
  {Cassidy}, \citenamefont {Schmits}, \citenamefont {Xu}, \citenamefont {Wang},
  \citenamefont {Krogstrup}, \citenamefont {Op~het Veld}, \citenamefont {Zuo},
  \citenamefont {Vos}, \citenamefont {Shen}, \citenamefont {Bouman},
  \citenamefont {Shojaei}, \citenamefont {Pennachio}, \citenamefont {Lee},
  \citenamefont {van Veldhoven}, \citenamefont {Koelling}, \citenamefont
  {Verheijen}, \citenamefont {Kouwenhoven}, \citenamefont {Palmstr{\o}m},\ and\
  \citenamefont {Bakkers}}]{Gazibegovic2017}%
  \BibitemOpen
  \bibfield  {author} {\bibinfo {author} {\bibfnamefont {S.}~\bibnamefont
  {Gazibegovic}}, \bibinfo {author} {\bibfnamefont {D.}~\bibnamefont {Car}},
  \bibinfo {author} {\bibfnamefont {H.}~\bibnamefont {Zhang}}, \bibinfo
  {author} {\bibfnamefont {S.~C.}\ \bibnamefont {Balk}}, \bibinfo {author}
  {\bibfnamefont {J.~A.}\ \bibnamefont {Logan}}, \bibinfo {author}
  {\bibfnamefont {M.~W.~A.}\ \bibnamefont {de~Moor}}, \bibinfo {author}
  {\bibfnamefont {M.~C.}\ \bibnamefont {Cassidy}}, \bibinfo {author}
  {\bibfnamefont {R.}~\bibnamefont {Schmits}}, \bibinfo {author} {\bibfnamefont
  {D.}~\bibnamefont {Xu}}, \bibinfo {author} {\bibfnamefont {G.}~\bibnamefont
  {Wang}}, \bibinfo {author} {\bibfnamefont {P.}~\bibnamefont {Krogstrup}},
  \bibinfo {author} {\bibfnamefont {R.~L.~M.}\ \bibnamefont {Op~het Veld}},
  \bibinfo {author} {\bibfnamefont {K.}~\bibnamefont {Zuo}}, \bibinfo {author}
  {\bibfnamefont {Y.}~\bibnamefont {Vos}}, \bibinfo {author} {\bibfnamefont
  {J.}~\bibnamefont {Shen}}, \bibinfo {author} {\bibfnamefont {D.}~\bibnamefont
  {Bouman}}, \bibinfo {author} {\bibfnamefont {B.}~\bibnamefont {Shojaei}},
  \bibinfo {author} {\bibfnamefont {D.}~\bibnamefont {Pennachio}}, \bibinfo
  {author} {\bibfnamefont {J.~S.}\ \bibnamefont {Lee}}, \bibinfo {author}
  {\bibfnamefont {P.~J.}\ \bibnamefont {van Veldhoven}}, \bibinfo {author}
  {\bibfnamefont {S.}~\bibnamefont {Koelling}}, \bibinfo {author}
  {\bibfnamefont {M.~A.}\ \bibnamefont {Verheijen}}, \bibinfo {author}
  {\bibfnamefont {L.~P.}\ \bibnamefont {Kouwenhoven}}, \bibinfo {author}
  {\bibfnamefont {C.~J.}\ \bibnamefont {Palmstr{\o}m}}, \ and\ \bibinfo
  {author} {\bibfnamefont {E.~P. A.~M.}\ \bibnamefont {Bakkers}},\ }\href
  {\doibase 10.1038/nature23468} {\bibfield  {journal} {\bibinfo  {journal}
  {Nature}\ }\textbf {\bibinfo {volume} {548}},\ \bibinfo {pages} {434}
  (\bibinfo {year} {2017})}\BibitemShut {NoStop}%
\bibitem [{\citenamefont {Ramos-Castillo}\ \emph {et~al.}(2015)\citenamefont
  {Ramos-Castillo}, \citenamefont {Reveles}, \citenamefont {Zope},\ and\
  \citenamefont {de~Coss}}]{Castillo2015}%
  \BibitemOpen
  \bibfield  {author} {\bibinfo {author} {\bibfnamefont {C.}~\bibnamefont
  {Ramos-Castillo}}, \bibinfo {author} {\bibfnamefont {J.}~\bibnamefont
  {Reveles}}, \bibinfo {author} {\bibfnamefont {R.}~\bibnamefont {Zope}}, \
  and\ \bibinfo {author} {\bibfnamefont {R.}~\bibnamefont {de~Coss}},\ }\href
  {\doibase 10.1021/acs.jpcc.5b02358} {\bibfield  {journal} {\bibinfo
  {journal} {The Journal of Physical Chemistry C}\ }\textbf {\bibinfo {volume}
  {119}},\ \bibinfo {pages} {8402} (\bibinfo {year} {2015})}\BibitemShut
  {NoStop}%
\bibitem [{\citenamefont {Koenig}\ \emph {et~al.}(2014)\citenamefont {Koenig},
  \citenamefont {Doganov}, \citenamefont {Schmidt}, \citenamefont
  {Castro~Neto},\ and\ \citenamefont {Özyilmaz}}]{Koenig2014}%
  \BibitemOpen
  \bibfield  {author} {\bibinfo {author} {\bibfnamefont {S.~P.}\ \bibnamefont
  {Koenig}}, \bibinfo {author} {\bibfnamefont {R.~A.}\ \bibnamefont {Doganov}},
  \bibinfo {author} {\bibfnamefont {H.}~\bibnamefont {Schmidt}}, \bibinfo
  {author} {\bibfnamefont {A.~H.}\ \bibnamefont {Castro~Neto}}, \ and\ \bibinfo
  {author} {\bibfnamefont {B.}~\bibnamefont {Özyilmaz}},\ }\href {\doibase
  10.1063/1.4868132} {\bibfield  {journal} {\bibinfo  {journal} {Applied
  Physics Letters}\ }\textbf {\bibinfo {volume} {104}},\ \bibinfo {pages}
  {103106} (\bibinfo {year} {2014})}\BibitemShut {NoStop}%
\bibitem [{\citenamefont {Li}\ \emph {et~al.}(2014)\citenamefont {Li},
  \citenamefont {Yu}, \citenamefont {Ye}, \citenamefont {Ge}, \citenamefont
  {Ou}, \citenamefont {Wu}, \citenamefont {Feng}, \citenamefont {Chen},\ and\
  \citenamefont {Zhang}}]{Li2014}%
  \BibitemOpen
  \bibfield  {author} {\bibinfo {author} {\bibfnamefont {L.}~\bibnamefont
  {Li}}, \bibinfo {author} {\bibfnamefont {Y.}~\bibnamefont {Yu}}, \bibinfo
  {author} {\bibfnamefont {G.~J.}\ \bibnamefont {Ye}}, \bibinfo {author}
  {\bibfnamefont {Q.}~\bibnamefont {Ge}}, \bibinfo {author} {\bibfnamefont
  {X.}~\bibnamefont {Ou}}, \bibinfo {author} {\bibfnamefont {H.}~\bibnamefont
  {Wu}}, \bibinfo {author} {\bibfnamefont {D.}~\bibnamefont {Feng}}, \bibinfo
  {author} {\bibfnamefont {X.~H.}\ \bibnamefont {Chen}}, \ and\ \bibinfo
  {author} {\bibfnamefont {Y.}~\bibnamefont {Zhang}},\ }\href {\doibase
  10.1038/nnano.2014.35} {\bibfield  {journal} {\bibinfo  {journal} {Nature
  Nanotechnology}\ }\textbf {\bibinfo {volume} {9}},\ \bibinfo {pages} {372}
  (\bibinfo {year} {2014})}\BibitemShut {NoStop}%
\bibitem [{\citenamefont {Son}(2007)}]{PhysRevB.75.235423}%
  \BibitemOpen
  \bibfield  {author} {\bibinfo {author} {\bibfnamefont {D.~T.}\ \bibnamefont
  {Son}},\ }\href {\doibase 10.1103/PhysRevB.75.235423} {\bibfield  {journal}
  {\bibinfo  {journal} {Phys. Rev. B}\ }\textbf {\bibinfo {volume} {75}},\
  \bibinfo {pages} {235423} (\bibinfo {year} {2007})}\BibitemShut {NoStop}%
\bibitem [{\citenamefont {Herbut}\ \emph {et~al.}(2008)\citenamefont {Herbut},
  \citenamefont {Juri\ifmmode \check{c}\else \v{c}\fi{}i\ifmmode~\acute{c}\else
  \'{c}\fi{}},\ and\ \citenamefont {Vafek}}]{PhysRevLett.100.046403}%
  \BibitemOpen
  \bibfield  {author} {\bibinfo {author} {\bibfnamefont {I.~F.}\ \bibnamefont
  {Herbut}}, \bibinfo {author} {\bibfnamefont {V.}~\bibnamefont {Juri\ifmmode
  \check{c}\else \v{c}\fi{}i\ifmmode~\acute{c}\else \'{c}\fi{}}}, \ and\
  \bibinfo {author} {\bibfnamefont {O.}~\bibnamefont {Vafek}},\ }\href
  {\doibase 10.1103/PhysRevLett.100.046403} {\bibfield  {journal} {\bibinfo
  {journal} {Phys. Rev. Lett.}\ }\textbf {\bibinfo {volume} {100}},\ \bibinfo
  {pages} {046403} (\bibinfo {year} {2008})}\BibitemShut {NoStop}%
\bibitem [{\citenamefont {Giannozzi}\ \emph {et~al.}(2009)\citenamefont
  {Giannozzi}, \citenamefont {Baroni}, \citenamefont {Bonini}, \citenamefont
  {Calandra}, \citenamefont {Car}, \citenamefont {Cavazzoni}, \citenamefont
  {Ceresoli}, \citenamefont {Chiarotti}, \citenamefont {Cococcioni},
  \citenamefont {Dabo}, \citenamefont {Corso}, \citenamefont {de~Gironcoli},
  \citenamefont {Fabris}, \citenamefont {Fratesi}, \citenamefont {Gebauer},
  \citenamefont {Gerstmann}, \citenamefont {Gougoussis}, \citenamefont
  {Kokalj}, \citenamefont {Lazzeri}, \citenamefont {Martin-Samos},
  \citenamefont {Marzari}, \citenamefont {Mauri}, \citenamefont {Mazzarello},
  \citenamefont {Paolini}, \citenamefont {Pasquarello}, \citenamefont
  {Paulatto}, \citenamefont {Sbraccia}, \citenamefont {Scandolo}, \citenamefont
  {Sclauzero}, \citenamefont {Seitsonen}, \citenamefont {Smogunov},
  \citenamefont {Umari},\ and\ \citenamefont {Wentzcovitch}}]{Giannozzi_2009}%
  \BibitemOpen
  \bibfield  {author} {\bibinfo {author} {\bibfnamefont {P.}~\bibnamefont
  {Giannozzi}}, \bibinfo {author} {\bibfnamefont {S.}~\bibnamefont {Baroni}},
  \bibinfo {author} {\bibfnamefont {N.}~\bibnamefont {Bonini}}, \bibinfo
  {author} {\bibfnamefont {M.}~\bibnamefont {Calandra}}, \bibinfo {author}
  {\bibfnamefont {R.}~\bibnamefont {Car}}, \bibinfo {author} {\bibfnamefont
  {C.}~\bibnamefont {Cavazzoni}}, \bibinfo {author} {\bibfnamefont
  {D.}~\bibnamefont {Ceresoli}}, \bibinfo {author} {\bibfnamefont {G.~L.}\
  \bibnamefont {Chiarotti}}, \bibinfo {author} {\bibfnamefont {M.}~\bibnamefont
  {Cococcioni}}, \bibinfo {author} {\bibfnamefont {I.}~\bibnamefont {Dabo}},
  \bibinfo {author} {\bibfnamefont {A.~D.}\ \bibnamefont {Corso}}, \bibinfo
  {author} {\bibfnamefont {S.}~\bibnamefont {de~Gironcoli}}, \bibinfo {author}
  {\bibfnamefont {S.}~\bibnamefont {Fabris}}, \bibinfo {author} {\bibfnamefont
  {G.}~\bibnamefont {Fratesi}}, \bibinfo {author} {\bibfnamefont
  {R.}~\bibnamefont {Gebauer}}, \bibinfo {author} {\bibfnamefont
  {U.}~\bibnamefont {Gerstmann}}, \bibinfo {author} {\bibfnamefont
  {C.}~\bibnamefont {Gougoussis}}, \bibinfo {author} {\bibfnamefont
  {A.}~\bibnamefont {Kokalj}}, \bibinfo {author} {\bibfnamefont
  {M.}~\bibnamefont {Lazzeri}}, \bibinfo {author} {\bibfnamefont
  {L.}~\bibnamefont {Martin-Samos}}, \bibinfo {author} {\bibfnamefont
  {N.}~\bibnamefont {Marzari}}, \bibinfo {author} {\bibfnamefont
  {F.}~\bibnamefont {Mauri}}, \bibinfo {author} {\bibfnamefont
  {R.}~\bibnamefont {Mazzarello}}, \bibinfo {author} {\bibfnamefont
  {S.}~\bibnamefont {Paolini}}, \bibinfo {author} {\bibfnamefont
  {A.}~\bibnamefont {Pasquarello}}, \bibinfo {author} {\bibfnamefont
  {L.}~\bibnamefont {Paulatto}}, \bibinfo {author} {\bibfnamefont
  {C.}~\bibnamefont {Sbraccia}}, \bibinfo {author} {\bibfnamefont
  {S.}~\bibnamefont {Scandolo}}, \bibinfo {author} {\bibfnamefont
  {G.}~\bibnamefont {Sclauzero}}, \bibinfo {author} {\bibfnamefont {A.~P.}\
  \bibnamefont {Seitsonen}}, \bibinfo {author} {\bibfnamefont {A.}~\bibnamefont
  {Smogunov}}, \bibinfo {author} {\bibfnamefont {P.}~\bibnamefont {Umari}}, \
  and\ \bibinfo {author} {\bibfnamefont {R.~M.}\ \bibnamefont {Wentzcovitch}},\
  }\href {\doibase 10.1088/0953-8984/21/39/395502} {\bibfield  {journal}
  {\bibinfo  {journal} {Journal of Physics: Condensed Matter}\ }\textbf
  {\bibinfo {volume} {21}},\ \bibinfo {pages} {395502} (\bibinfo {year}
  {2009})}\BibitemShut {NoStop}%
\bibitem [{\citenamefont {Perdew}\ \emph {et~al.}(1996)\citenamefont {Perdew},
  \citenamefont {Burke},\ and\ \citenamefont
  {Ernzerhof}}]{PhysRevLett.77.3865}%
  \BibitemOpen
  \bibfield  {author} {\bibinfo {author} {\bibfnamefont {J.~P.}\ \bibnamefont
  {Perdew}}, \bibinfo {author} {\bibfnamefont {K.}~\bibnamefont {Burke}}, \
  and\ \bibinfo {author} {\bibfnamefont {M.}~\bibnamefont {Ernzerhof}},\ }\href
  {\doibase 10.1103/PhysRevLett.77.3865} {\bibfield  {journal} {\bibinfo
  {journal} {Phys. Rev. Lett.}\ }\textbf {\bibinfo {volume} {77}},\ \bibinfo
  {pages} {3865} (\bibinfo {year} {1996})}\BibitemShut {NoStop}%
\bibitem [{\citenamefont {Monkhorst}\ and\ \citenamefont
  {Pack}(1976)}]{PhysRevB.13.5188}%
  \BibitemOpen
  \bibfield  {author} {\bibinfo {author} {\bibfnamefont {H.~J.}\ \bibnamefont
  {Monkhorst}}\ and\ \bibinfo {author} {\bibfnamefont {J.~D.}\ \bibnamefont
  {Pack}},\ }\href {\doibase 10.1103/PhysRevB.13.5188} {\bibfield  {journal}
  {\bibinfo  {journal} {Phys. Rev. B}\ }\textbf {\bibinfo {volume} {13}},\
  \bibinfo {pages} {5188} (\bibinfo {year} {1976})}\BibitemShut {NoStop}%
\bibitem [{\citenamefont {Janssen}\ and\ \citenamefont
  {Vojta}(2019)}]{Janssen2019}%
  \BibitemOpen
  \bibfield  {author} {\bibinfo {author} {\bibfnamefont {L.}~\bibnamefont
  {Janssen}}\ and\ \bibinfo {author} {\bibfnamefont {M.}~\bibnamefont
  {Vojta}},\ }\href {\doibase 10.1088/1361-648x/ab283e} {\bibfield  {journal}
  {\bibinfo  {journal} {Journal of Physics: Condensed Matter}\ }\textbf
  {\bibinfo {volume} {31}},\ \bibinfo {pages} {423002} (\bibinfo {year}
  {2019})}\BibitemShut {NoStop}%
\bibitem [{\citenamefont {Parameswaran}\ and\ \citenamefont
  {Feldman}(2019)}]{Parameswaran2019}%
  \BibitemOpen
  \bibfield  {author} {\bibinfo {author} {\bibfnamefont {S.~A.}\ \bibnamefont
  {Parameswaran}}\ and\ \bibinfo {author} {\bibfnamefont {B.~E.}\ \bibnamefont
  {Feldman}},\ }\href {\doibase 10.1088/1361-648x/ab0636} {\bibfield  {journal}
  {\bibinfo  {journal} {Journal of Physics: Condensed Matter}\ }\textbf
  {\bibinfo {volume} {31}},\ \bibinfo {pages} {273001} (\bibinfo {year}
  {2019})}\BibitemShut {NoStop}%
\bibitem [{\citenamefont {Sevin{\c{c}}li}\ \emph {et~al.}(2019)\citenamefont
  {Sevin{\c{c}}li}, \citenamefont {Roche}, \citenamefont {Cuniberti},
  \citenamefont {Brandbyge}, \citenamefont {Gutierrez},\ and\ \citenamefont
  {Sandonas}}]{Sevinli2019}%
  \BibitemOpen
  \bibfield  {author} {\bibinfo {author} {\bibfnamefont {H.}~\bibnamefont
  {Sevin{\c{c}}li}}, \bibinfo {author} {\bibfnamefont {S.}~\bibnamefont
  {Roche}}, \bibinfo {author} {\bibfnamefont {G.}~\bibnamefont {Cuniberti}},
  \bibinfo {author} {\bibfnamefont {M.}~\bibnamefont {Brandbyge}}, \bibinfo
  {author} {\bibfnamefont {R.}~\bibnamefont {Gutierrez}}, \ and\ \bibinfo
  {author} {\bibfnamefont {L.~M.}\ \bibnamefont {Sandonas}},\ }\href {\doibase
  10.1088/1361-648x/ab119a} {\bibfield  {journal} {\bibinfo  {journal} {Journal
  of Physics: Condensed Matter}\ }\textbf {\bibinfo {volume} {31}},\ \bibinfo
  {pages} {273003} (\bibinfo {year} {2019})}\BibitemShut {NoStop}%
\bibitem [{\citenamefont {Herrero}\ and\ \citenamefont
  {Ram{\'{\i}}rez}(2014)}]{Herrero2014}%
  \BibitemOpen
  \bibfield  {author} {\bibinfo {author} {\bibfnamefont {C.~P.}\ \bibnamefont
  {Herrero}}\ and\ \bibinfo {author} {\bibfnamefont {R.}~\bibnamefont
  {Ram{\'{\i}}rez}},\ }\href {\doibase 10.1088/0953-8984/26/23/233201}
  {\bibfield  {journal} {\bibinfo  {journal} {Journal of Physics: Condensed
  Matter}\ }\textbf {\bibinfo {volume} {26}},\ \bibinfo {pages} {233201}
  (\bibinfo {year} {2014})}\BibitemShut {NoStop}%
\bibitem [{\citenamefont {Zaitsev}\ \emph {et~al.}(2018)\citenamefont
  {Zaitsev}, \citenamefont {Jakob},\ and\ \citenamefont
  {Tonner}}]{Zaitsev2018}%
  \BibitemOpen
  \bibfield  {author} {\bibinfo {author} {\bibfnamefont {N.~L.}\ \bibnamefont
  {Zaitsev}}, \bibinfo {author} {\bibfnamefont {P.}~\bibnamefont {Jakob}}, \
  and\ \bibinfo {author} {\bibfnamefont {R.}~\bibnamefont {Tonner}},\ }\href
  {\doibase 10.1088/1361-648x/aad576} {\bibfield  {journal} {\bibinfo
  {journal} {Journal of Physics: Condensed Matter}\ }\textbf {\bibinfo {volume}
  {30}},\ \bibinfo {pages} {354001} (\bibinfo {year} {2018})}\BibitemShut
  {NoStop}%
\bibitem [{\citenamefont {Mills}\ and\ \citenamefont {J\'onsson}(1994)}]{16}%
  \BibitemOpen
  \bibfield  {author} {\bibinfo {author} {\bibfnamefont {G.}~\bibnamefont
  {Mills}}\ and\ \bibinfo {author} {\bibfnamefont {H.}~\bibnamefont
  {J\'onsson}},\ }\href {\doibase 10.1103/PhysRevLett.72.1124} {\bibfield
  {journal} {\bibinfo  {journal} {Phys. Rev. Lett.}\ }\textbf {\bibinfo
  {volume} {72}},\ \bibinfo {pages} {1124} (\bibinfo {year}
  {1994})}\BibitemShut {NoStop}%
\bibitem [{\citenamefont {Castro~Neto}\ \emph
  {et~al.}(2009{\natexlab{b}})\citenamefont {Castro~Neto}, \citenamefont
  {Guinea}, \citenamefont {Peres}, \citenamefont {Novoselov},\ and\
  \citenamefont {Geim}}]{17}%
  \BibitemOpen
  \bibfield  {author} {\bibinfo {author} {\bibfnamefont {A.~H.}\ \bibnamefont
  {Castro~Neto}}, \bibinfo {author} {\bibfnamefont {F.}~\bibnamefont {Guinea}},
  \bibinfo {author} {\bibfnamefont {N.~M.~R.}\ \bibnamefont {Peres}}, \bibinfo
  {author} {\bibfnamefont {K.~S.}\ \bibnamefont {Novoselov}}, \ and\ \bibinfo
  {author} {\bibfnamefont {A.~K.}\ \bibnamefont {Geim}},\ }\href {\doibase
  10.1103/RevModPhys.81.109} {\bibfield  {journal} {\bibinfo  {journal} {Rev.
  Mod. Phys.}\ }\textbf {\bibinfo {volume} {81}},\ \bibinfo {pages} {109}
  (\bibinfo {year} {2009}{\natexlab{b}})}\BibitemShut {NoStop}%
\bibitem [{\citenamefont {Han}\ \emph {et~al.}(2007)\citenamefont {Han},
  \citenamefont {\"Ozyilmaz}, \citenamefont {Zhang},\ and\ \citenamefont
  {Kim}}]{18}%
  \BibitemOpen
  \bibfield  {author} {\bibinfo {author} {\bibfnamefont {M.~Y.}\ \bibnamefont
  {Han}}, \bibinfo {author} {\bibfnamefont {B.}~\bibnamefont {\"Ozyilmaz}},
  \bibinfo {author} {\bibfnamefont {Y.}~\bibnamefont {Zhang}}, \ and\ \bibinfo
  {author} {\bibfnamefont {P.}~\bibnamefont {Kim}},\ }\href {\doibase
  10.1103/PhysRevLett.98.206805} {\bibfield  {journal} {\bibinfo  {journal}
  {Phys. Rev. Lett.}\ }\textbf {\bibinfo {volume} {98}},\ \bibinfo {pages}
  {206805} (\bibinfo {year} {2007})}\BibitemShut {NoStop}%
\bibitem [{\citenamefont {Mermin}\ and\ \citenamefont {Wagner}(1966)}]{20}%
  \BibitemOpen
  \bibfield  {author} {\bibinfo {author} {\bibfnamefont {N.~D.}\ \bibnamefont
  {Mermin}}\ and\ \bibinfo {author} {\bibfnamefont {H.}~\bibnamefont
  {Wagner}},\ }\href {\doibase 10.1103/PhysRevLett.17.1133} {\bibfield
  {journal} {\bibinfo  {journal} {Phys. Rev. Lett.}\ }\textbf {\bibinfo
  {volume} {17}},\ \bibinfo {pages} {1133} (\bibinfo {year}
  {1966})}\BibitemShut {NoStop}%
\bibitem [{\citenamefont {Belitz}\ \emph {et~al.}(2005)\citenamefont {Belitz},
  \citenamefont {Kirkpatrick},\ and\ \citenamefont
  {Vojta}}]{RevModPhys.77.579}%
  \BibitemOpen
  \bibfield  {author} {\bibinfo {author} {\bibfnamefont {D.}~\bibnamefont
  {Belitz}}, \bibinfo {author} {\bibfnamefont {T.~R.}\ \bibnamefont
  {Kirkpatrick}}, \ and\ \bibinfo {author} {\bibfnamefont {T.}~\bibnamefont
  {Vojta}},\ }\href {\doibase 10.1103/RevModPhys.77.579} {\bibfield  {journal}
  {\bibinfo  {journal} {Rev. Mod. Phys.}\ }\textbf {\bibinfo {volume} {77}},\
  \bibinfo {pages} {579} (\bibinfo {year} {2005})}\BibitemShut {NoStop}%
\bibitem [{\citenamefont {H\"ogl}\ \emph {et~al.}(2020)\citenamefont {H\"ogl},
  \citenamefont {Frank}, \citenamefont {Kochan}, \citenamefont {Gmitra},\ and\
  \citenamefont {Fabian}}]{PhysRevB.101.245441}%
  \BibitemOpen
  \bibfield  {author} {\bibinfo {author} {\bibfnamefont {P.}~\bibnamefont
  {H\"ogl}}, \bibinfo {author} {\bibfnamefont {T.}~\bibnamefont {Frank}},
  \bibinfo {author} {\bibfnamefont {D.}~\bibnamefont {Kochan}}, \bibinfo
  {author} {\bibfnamefont {M.}~\bibnamefont {Gmitra}}, \ and\ \bibinfo {author}
  {\bibfnamefont {J.}~\bibnamefont {Fabian}},\ }\href {\doibase
  10.1103/PhysRevB.101.245441} {\bibfield  {journal} {\bibinfo  {journal}
  {Phys. Rev. B}\ }\textbf {\bibinfo {volume} {101}},\ \bibinfo {pages}
  {245441} (\bibinfo {year} {2020})}\BibitemShut {NoStop}%
\end{thebibliography}%
\bibliographystyle{apsrev4-1}

\onecolumngrid
\section*{SUPPLEMENTARY MATERIAL}

\onecolumngrid
Rational control of individual molecular spins in nanoelectronics devices is a pivotal prerequisite to fulfill its potential promised by molecular spintronics\cite{PhysRevB.89.201110,PhysRevB.75.235423}. There is a rapidly growing interest in the transmission of information encoded by electron and atomic spins which, in principle, can be realized by the propagation of spin waves or magnetization waves through low-dimensional sheets. In this aspect, weak-coupling phenomena of the two-dimensional Hubbard model is gaining momentum as a new interesting research field recently because of its extraordinarily rich behavior as a function of the carrier density and model parameters\cite{PhysRevLett.99.226803,PhysRevLett.100.046403}. \\

In this regard, establishing an effective \textit{ab initio} Schrodinger equation for spin waves incorporating analytic response functions with suitable boundary conditions for spin wave controlled transmission is absolute requirement to validate the theoretical understanding about antiferromagnetic spin-wave field-effect transistor\cite{PhysRevLett.107.196803,PhysRevLett.108.046602,PhysRevB.86.165127}. In this aspect, this report is the first of its kind where we propose that the spin wave assisted field-effect transmission may be selectively used for controlling individual molecules\cite{PhysRevB.87.205138} embedded in the dual-gate field-effect transistor architecture.\\


\
\onecolumngrid
\subsection*{Free susceptibility}
\begin{figure}[th!]     
\centering           
\includegraphics[width=7.5cm,height=5.5cm]{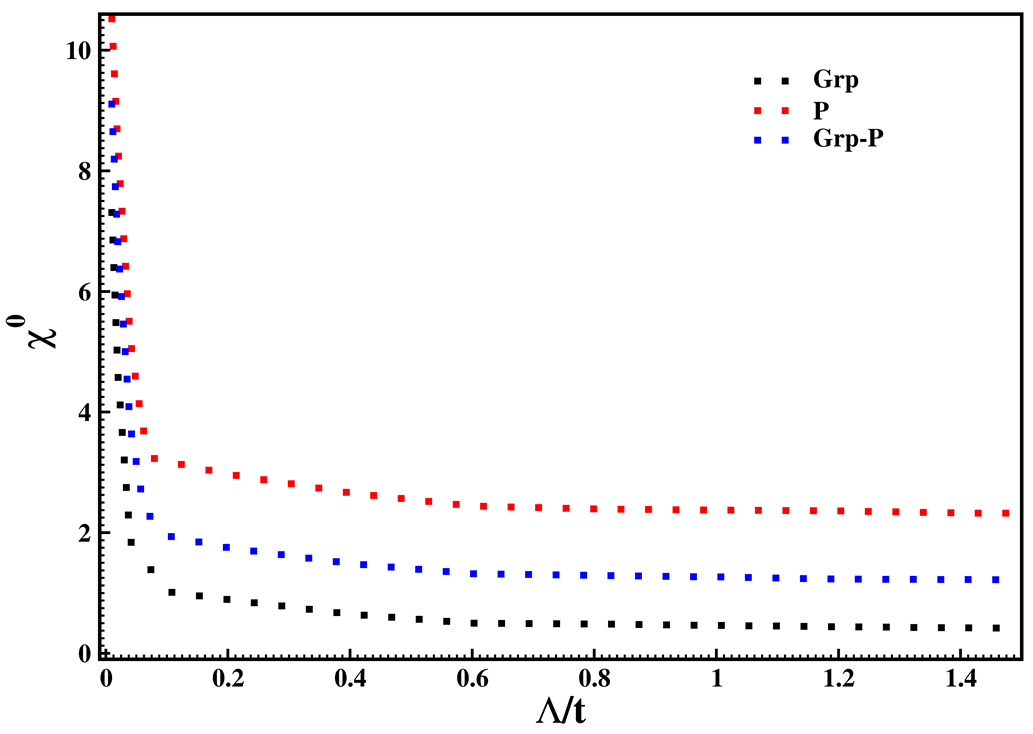}
\caption{Free (non-interacting) susceptibilities of {\it Grp-P} heterostructure system with a comparison to the pristine graphene and phosphorene sheet for t=0 and $|\mu|$=0.005.}
   \label{fig_sm_1}
\end{figure}

Moreover, we have shown the trend of free susceptibilities ({$\chi$}$^{0}$) at specific choice of {\it t} and $\mu$ values (shown in Fig. ~\ref{fig_sm_1}), strongly supporting the trend of homogeneous susceptibility. 

\onecolumngrid
\subsection*{Phase Transition}
Let us consider the phase transition in the weakly coupled heterostructure due to the instabilities prevailing on the surface.  
\begin{figure}[th!]     
\centering           
\includegraphics[width=6.5cm,height=8.5cm]{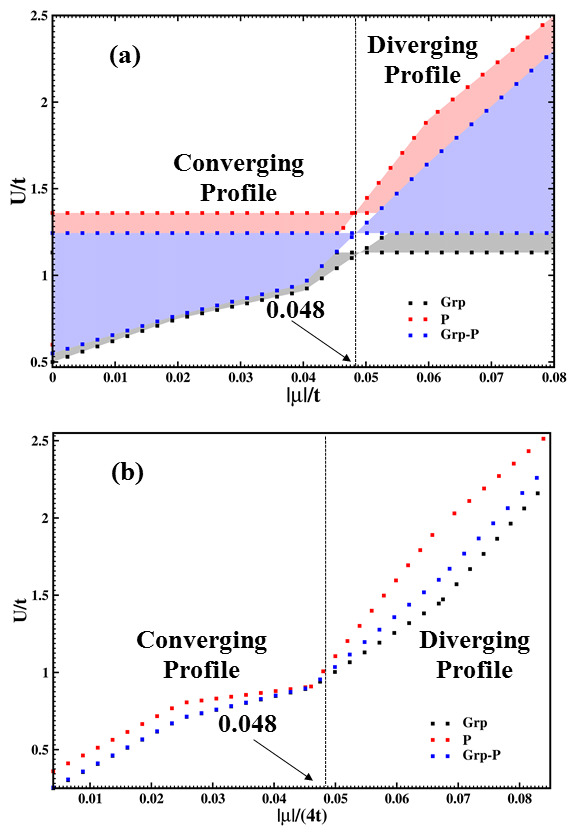}
\caption{Spin-wave density dependent phase diagram for (a) The ($\mu$, U) phase diagram for t=0 close to half-filling (n$\approx$1, $|\mu|\rightarrow$0) region; the solid line represents the spin-density-wave regime of the {\it Grp-P} heterostructure system with a comparison to the pristine graphene and phosphorene sheet. Similarly, (b) The (t, U) phase diagram for 4t$<$0 value below the half-filling ($|\mu|<$0) region; the dotted lines represent the spin-density-wave regime of the same three systems. }
 \label{fig_sm_2}
\end{figure}

We show the phase diagram of $\mu$ and {\it U} at half-filling measured by the prevailing uncertainty from the flow equation (shown in Fig.~\ref{fig_sm_2}(a)). The leading uncertainty region in case of {\it commensurate} spin-wave density is detached from the {\it s}-wave coupling region {\it via} a narrow strip where {\it incommensurate} spin-wave density fluctuations show dominant effect. For small value of {\it U}, the region surrounding half-filling is exponentially small, where uncertainties in spin-wave density play a dominant effect. The value of ({\it t, U}), {\it in-plane} phase diagram is shown in Fig.~\ref{fig_sm_2}(b) with $|\mu|$=4t and $\mu${$<$}0 (below half-filling).

\onecolumngrid
In this case, all the steps are required for a clear udnerstanding of the facts. As a result, rational control of individual molecular spins in nanoelectronics devices is a pivotal prerequisite to fulfill its potential applicability. Meanwhile, there is a rapidly growing interest in the transmission of information encoded by electron and atomic spins which, in principle, can be realized by the propagation of spin waves or magnetization waves through low-dimensional systems. In this aspect, weak-coupling phenomena of the two-dimensional Hubbard model is gaining momentum as a new interesting research field recently because of its extraordinarily rich behavior as a function of the carrier density and model parameters. We show that this finding supports the quantitative realization of experimental and theoretical systems farther from their critical points before the underlying field theory is well understood for potential application.  

\onecolumngrid
\subsection*{First Principle based DFT Calculations}
QUANTUM Espresso codes~\cite{Giannozzi_2009} based on Perdew-Burke-Ernzerhof (PBE) with generalized gradient approximation (GGA) is used in our DFT calculations \cite{PhysRevLett.77.3865} in the framework of local spin density approximation (LSDA) {\it i.e.} all calculations are carried out taking into account the presence of electron spins with time-reversal (TR) symmetry breaking along {\it z}-axis. Supercells having larger than 20 $\AA{}$ lattices are considered for first principle calculations to avoid periodic images. Brillouin zone integration has been performed with a 3$\times$3$\times$1 k points grid sampling at an energy cutoff of 60 Ry based on the Monkhorst and Pack scheme \cite{PhysRevB.13.5188}. Densities of states have been obtained on a {\it k}-point sampling of 9$\times$9$\times$1 with smearing effect at 540 Ry as charge density cut-off energy. The supercell structure has been optimized till the force and energy per each atom attains the value within 0.01 eV/$\AA{}$ and 2.0$\times$10$^{-7}$ eV, respectively. Three types of systems, such as graphene monolayer, phosphorene monolayer and graphene/phosphorene bilayer have been considered in our calculations. The bilayer heterostructure has been made of the host ({\it i.e.} phosphorene) on graphene substrate ({\it i.e.} bilayer of graphene/phosphorene).  In this case, the supercell, 4$\sqrt{3}\times$4$\sqrt{3}$ graphene on $\sqrt{27}\times\sqrt{27}$ BP with a rotation angle of 12$^{0}$ was chosen with 48 Carbon atoms in graphene layer. The lattice mismatch is less than 0.24~$\%$. In case of monolayer systems, same number of atoms are taken in the simulation cell.  

\nocite{*}

\end{document}